\documentclass[reprint,aps,pre,floatfix]{revtex4-1}
\usepackage{xcolor}
\usepackage[normalem]{ulem}
\usepackage{amsmath,amsfonts,amscd,amssymb,graphicx}
\usepackage{lineno}
\usepackage{dcolumn}% Align table columns on decimal point
\usepackage{overpic}
\usepackage{color}
\usepackage{placeins}
\usepackage{graphicx}
\newcommand{\bX}{\mathbf{X}}
\newcommand{\bx}{\mathbf{x}}
\newcommand{\by}{\mathbf{y}}
\newcommand{\bA}{\mathbf{A}}
\newcommand{\bB}{\mathbf{B}}
\newcommand{\bC}{\mathbf{C}}
\newcommand{\bu}{\mathbf{u}}
\newcommand{\bU}{\mathbf{U}}
\newcommand{\bh}{\mathbf{h}}

\DeclareMathOperator*{\argmin}{arg\,min}
\usepackage{mathrsfs}
\usepackage{subcaption}
\usepackage{adjustbox}
\usepackage{rotating}
\usepackage{float}
\usepackage{lipsum}
\makeatletter
\let\newfloat\newfloat@ltx
\makeatother
\usepackage{algorithm}
\usepackage{algpseudocode}
\algnewcommand\algorithmicinput{\textbf{Input:}}
\algnewcommand\Input{\item[\algorithmicinput]}

\begin{document}

\title{Reservoir computing for system identification and predictive control with limited data}

\author{Jan P. Williams$^*$, J. Nathan Kutz$^{\dag}$ and Krithika Manohar$^*$}
\affiliation{$^*$Department of Mechanical Engineering, University of Washington, Seattle, WA} 
\affiliation{$^\dag$  Departments of Applied Mathematics and Electrical and Computer Engineering, University of Washington, Seattle, WA}
 
\begin{abstract}
Model predictive control (MPC) is an industry standard control technique that iteratively solves an open-loop optimization problem to guide a system towards a desired state or trajectory. Consequently, an accurate forward model of system dynamics is critical for the efficacy of MPC and much recent work has been aimed at the use of neural networks to act as data-driven surrogate models to enable MPC. Perhaps the most common network architecture applied to this task is the recurrent neural network (RNN) due to its natural interpretation as a dynamical system. In this work, we assess the ability of RNN variants to both learn the dynamics of benchmark control systems and serve as surrogate models for MPC. We find that echo state networks (ESNs) have a variety of benefits over competing architectures, namely reductions in computational complexity, longer valid prediction times, and reductions in cost of the MPC objective function.
\end{abstract}

\maketitle

\section{Introduction}
Model Predictive Control (MPC) is a widely used control algorithm known for its flexibility in the optimal control of complex dynamical systems \cite{kouvaritakis_model_nodate, schwenzer_review_2021}. It allows for the prediction of future system behavior by optimizing control actions over a specified time horizon, using a mathematical model of the underlying dynamical system. Since its introduction in the 1980s, MPC has been widely adopted due to its ability to control multi-input multi-output (MIMO) systems, as well as strongly nonlinear dynamics. On the other hand, the computational cost of repeated online optimization can render the method infeasible for dynamical systems with short characteristic timescales or computationally expensive forward models. Moreover, we are increasingly presented with dynamical systems for which we have no high-fidelity model derived from first principles. As a result, recent work has aimed at developing data-driven surrogate models that can act in the place of a derived model in MPC \cite{wang_fully_2006, morton_deep_2018, bieker_deep_2020, lee_identification_2000, terzi_learning_2021, jeon_lstm-based_2021, huang_lstm-mpc_2023, pan_model_2012, draeger_model_1995, jordanou_nonlinear_2018, bonassi_nonlinear_2024, bonassi_recurrent_2022, kaiser_sparse_2018}. In particular, variants of recurrent neural networks (RNNs) have been demonstrated to be effective across a range of benchmark control systems. This work provides an in-depth empirical comparison of competing RNN architectures for use as surrogate models with MPC. 

RNNs are a class of artificial neural networks designed specifically to process sequential data and have shown remarkable success in applications ranging from natural language processing to time series forecasting \cite{rumelhart_learning_1987, hochreiter_long_1997, hewamalage_recurrent_2021}. RNNs propagate an internal latent state sequentially, allowing them to seamlessly incorporate information from many previous time steps. This makes RNNs particularly well-suited for modeling dynamical systems where the current state of the system depends on the previous states. Indeed, the natural interpretation of a trained RNN is as a dynamical system, while the action of many other neural networks is naturally interpreted as a function.

The integration of RNNs into MPC offers several potential benefits. First, RNN-based surrogate models can significantly reduce the computational load associated with solving the optimization problem in MPC. This is achieved by providing rapid predictions of the system's future states, thereby enabling faster solution times. Second, RNNs can handle nonlinearities and complex interactions within the system that might be challenging to capture with traditional modeling approaches. This flexibility can allow for more accurate and robust control, particularly in systems with strongly nonlinear behaviors or where detailed first-principle models are unavailable or impractical to develop.\begin{figure*}[!ht]
    \centering
    \begin{overpic}[width=0.90\textwidth]{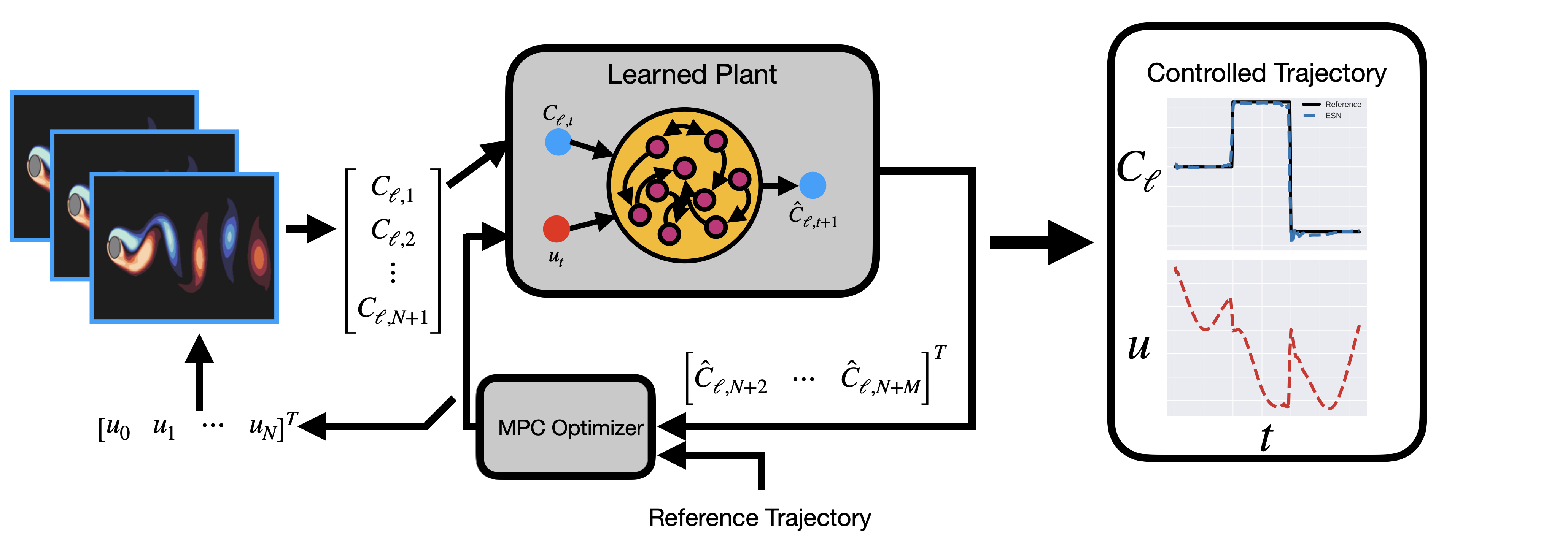} 
    \end{overpic}
    \caption{Summary diagram of echo state network (ESN) based model predictive control (MPC). An ESN trained to model plant dynamics acts as a surrogate model capable of rapidly forecasting the system under given control inputs. Online optimizations compute a sequence of control actions that minimize the deviation of the plant from a reference trajectory. }
    \label{fig:overview}
\end{figure*}

The efficacy of RNNs in modeling dynamical systems is demonstrated by their strong performance in forecasting chaotic systems. Variants of RNNs have achieved state-of-the-art forecasting performance in complex systems such as the Kuramoto Sivashinsky system, while retaining the long-term ergodic properties and Lyapunov spectrum of the underlying dynamics  \cite{platt_systematic_2022, vlachas_data-driven_2018, chattopadhyay_data-driven_2020}. Vlachas et al. \cite{vlachas_backpropagation_2020} provide an in-depth comparison of recurrent architectures trained through backpropagation through time (BPTT) and those that are a type of reservoir computer (RC). However, the body of literature exploring RNNs ability to replicate the climate of chaotic systems primarily considers autonomous dynamical systems, while MPC must handle non-autonomous dynamics. Furthermore, the computational requirements of an RNN surrogate model used with MPC are stricter than required for autonomous forecasts alone. 

Existing work has shown that both BPTT and RC based surrogate models can achieve satisfactory control performance in a range of systems \cite{wang_fully_2006, huang_lstm-mpc_2023, jeon_lstm-based_2021, jordanou_nonlinear_2018, armenio_model_2019}. Among these, RNNs have been successfully applied to control chaotic, fluidic systems \cite{bieker_deep_2020}. However, there exist few direct comparisons between BPTT and RC based approaches. Comparisons across separate works is difficult as a result of the adoption of different evaluation metrics, control parameters, and example systems. In this work, we provide a standardized comparison of the training protocols and control performance of long-short term memory (LSTM) networks, gated recurrent units (GRU), and echo state networks (ESN). LSTMs and GRUs are variants of RNNs that rely on gating mechanisms and backpropagation through time for training while ESNs are a form of reservoir computer. Additionally, comparisons to fully-connected networks (FCN) \cite{draeger_model_1995} and dynamic mode decomposition with control (DMDc) \cite{proctor_dynamic_2016} approaches are provided. % For this reason, we defer the discussion and presentation of GRUs to the SI. 

We demonstrate that ESNs are particularly suitable surrogate models for use with model predictive control. Across all considered benchmark systems exhibiting a range of dynamical behaviors, ESNs provide the most accurate open loop approximation of the dynamics. In turn, the closed loop performance of ESN based MPC is superior to that of the considered gated recurrent architectures. These results also hold in the presence of additive noise and with limited training data. Moreover, the use of ESNs instead of architectures trained via BPTT allows for drastic reductions in training time.

\section{Background}
Model predictive control (MPC) is the iterative optimization of a finite-horizon, open-loop optimal control problem, requiring re-optimization for new time horizons. Typically, MPC assumes a discrete-time model of system dynamics, $\hat \bx _{j+1} = f(\bx _j, \bu _j)$ where $\bx_j$ is the system state at time $j$ and $\bu _j$ is the exogenous control signal. Redefining $t=0$ for each optimization, a cost function is posed of the form
\begin{align}
    J(U, t) &= \sum _{j=1} ^T \epsilon _j^T(f) Q_1 \epsilon _j(f) + \bu _{j}^T Q_2 \bu _{j} + \Delta \bu _{j}^T Q_3 \Delta \bu _{j} \notag\\
    \epsilon _j (f) &= \mathbf{r}_j - f(\hat \bx _{j-1}, \bu _{j-1}) \label{eq:mpc_cost}
\end{align} % Second line no number
where $U = \{\bu _{t+1}, \dots, \bu _{t+M}\}$, $\mathbf r _j$ is the desired system state at time $j$, and $\Delta \bu_j = \bu_j - \bu _{j-1}$. $Q_1$ is a positive semi-definite weight matrix penalizing the deviation from the desired trajectory, and $Q_2, Q_3$ are positive definite matrices that penalize the magnitude of the control input and rapid control fluctuations, respectively. $T$ is the finite forecast horizon of the model predictive controller. $M$ is the number of control actions over which to optimize and must satisfy $M \leq T.$  After the $M$-th step of the forecast, control actions are frequently set to a fixed value. At time $t$, a model predictive controller must compute
\begin{equation}
    U_* = \argmin _{U \in \mathcal B} J(U, t)
    \label{eq:mpc_minimization}
\end{equation}
where $\mathcal B$ defines the set of feasible control actions.

The requirements of a surrogate model to be used with MPC are twofold. First, the model must be capable of generating accurate forecasts of the control system given a sequence of control inputs. Second, the model must provide computationally cheap predictions to allow for the online minimization of the MPC objective function. In this work, we consider a total of five candidate architectures for data-driven surrogate modeling. Each model is trained to generate one-step forecasts which can subsequently be used to forecast arbitrarily many timesteps ahead. 

% Perhaps the simplest forward dynamical model one could learn from data is a function $f$ parameterized by $\theta$ and of the form 
% \begin{equation}
%     \mathbf{x}_{j+1} = f_\theta \left( \mathbf{x}_{j}, \mathbf{u}_{j} \right). \label{eq:ffm_notds}
% \end{equation}
% The straightforward extension of this model to incorporate time-delays is given by 
% \begin{equation}
%     \mathbf{x}_{j+1} = f_\theta \left( \mathbf{x}_{j}, \dots, \mathbf{x}_{j-k}, \mathbf{u}_{j}, \dots, \mathbf{u}_{j-k}\right), 
%     \label{eq:ffm}
% \end{equation}
% where $k$ is the maximal time-delay included in the model. The training protocol for $f_\theta$ will of course vary depending on the selected structure of $f_\theta.$ In this subsection, we consider two candidate forms of $f_\theta$, namely a linear model and a fully-connected neural network.
\subsection{Linear System Identification}
The control of linear time-invariant (LTI) systems has been studied for over a century and there exist a wide variety of control techniques when such a model is available, e.g. linear quadratic regulators, linear quadratic Gaussian regulators, PID control, etc. For discrete-time systems, such methods assume a system of the form 
\begin{align}
    \bx _{j+1} &= \bA \bx _{j}+ \bB \bu _{j} \notag \\
    \by _{j} &= \bC \bx_{j},
    \label{eq:lti}
\end{align}
where $\bx _{j}$ represents the system state at time $j$, $\by _{j}$ denotes the measurement the corresponding measurement of the system, and $\bu_{j}$ represents the exogenous control input of the system. Because the control of such systems is well-understood and the model is interpretable, constructing a data-driven model of the form Eq. \ref{eq:lti} has been an attractive paradigm for system identification. In the simplest case when full-state measurements are available, i.e. $\bC = \mathbb I$, the identity matrix, one assumes access to a set of training data pairs $(\bx_{j}, \bu_{j})$ for $0 \leq j \leq N.$ From this training data, we construct
\begin{align}
    \bX &= \begin{bmatrix}
        \bx _{0} & \bx _{1} & \cdots &\bx _{N-1}
    \end{bmatrix}, \notag \\
    \bX ' &= \begin{bmatrix}
        \bx _1 & \bx _2 & \cdots &\bx _N
    \end{bmatrix}, \notag \\
    \bU &= \begin{bmatrix}
        \bu _{0} & \bu_1 & \cdots & \bu_{N-1}
    \end{bmatrix}.
\end{align}
The matrices $\bA$ and $\bB$ can then be estimated using linear regression 
\begin{equation}
     \argmin _{\Theta} \| \begin{bmatrix}
         \bX ^T & \bU ^T
     \end{bmatrix} \Theta - (\bX ')^T \|^2_2,
\end{equation}
for the parameters $\Theta = \begin{bmatrix}
    \bA & \bB
\end{bmatrix}^T$.  In the idealized setting of an LTI system with nondegenerate eigenvalues and zero noise, the above procedure will exactly identify $\bA$ and $\bB$. 

A similar result can be obtained when $\bC \neq \mathbb I$ through the incorporation of time-delays in the model, provided the resulting system is observable. This comes from the fact that an $n$-th order ODE can be rewritten as a system of $n$ first order ODEs and vice versa. Let $k$ denote the number of included time-delays and
\begin{align}
    \mathbf{Y} &= \begin{bmatrix}
        \mathbf{y}_k & \mathbf{y}_{k+1} & \cdots & \mathbf{y}_{N-1}\\
        \vdots & \vdots & \ddots & \vdots \\
        \mathbf{y}_0 & \mathbf{y}_1 & \cdots & \mathbf{y}_{N-k}
    \end{bmatrix} \notag\\
    \mathbf{Y}' &= \begin{bmatrix}
        \mathbf{y}_{k+1} & \mathbf{y}_{k+2} & \cdots & \mathbf{y}_{N}\\
        \vdots & \vdots & \ddots & \vdots \\
        \mathbf{y}_1 & \mathbf{y}_2 & \cdots & \mathbf{y}_{N-k+1}.
    \end{bmatrix}.
\end{align}
As before, we learn a dynamical operator, $\Theta$, using linear regression,
\begin{equation}
    \argmin _{\Theta} \| \begin{bmatrix}
        \mathbf{Y} ^T & \mathbf{U}^ T
    \end{bmatrix} \Theta - (\mathbf{Y}' )^T \|^2_2.
\end{equation}
This formulation is equivalent to dynamic mode decomposition with control (DMDc) \cite{proctor_dynamic_2016} assuming no rank truncation. Such DMDc models can be directly used for MPC, but they are limited by the inherent assumption of linearity. Highly nonlinear systems frequently require more complex methods for satisfactory MPC performance. 

\subsection{Neural Network Models}
Fully-connected neural networks (FCN) offer increased expressiveness in comparison to linear models at the cost of interpretability. FCNs can be viewed as a series of matrix multiplications punctuated by nonlinear activation functions. Concretely, a two-layer fully connected network $f_\theta$ can be expressed as 
\begin{equation}
    f_\theta (\mathbf x^{k}_{j}, \mathbf u^k_{j}) = W_1\left(R\left(W_0 \begin{bmatrix}
        \bx ^k_{j}\\
        \bu ^k_{j}
    \end{bmatrix} + b_0 \right) \right) + b_1,
    \label{eq:fcn}
\end{equation}
where $\theta = \{W_1, W_2, b_1, b_2\}$ are trainable parameters and $\bx ^k _{j}$, $\bu ^k_{j}$ are the state and control vectors at time $j$ augmented by up to $k$ time-delays. $R$ is a nonlinear activation function that operates elementwise. We seek $\theta$ such that the MSE of a one-step forecast is minimized over a set of $N+1$ training states and control inputs,
\begin{equation}
    f_\theta = \arg \min _\theta \frac{1}{N - k} \sum _{i=k}^{N} \|f_\theta \left( \mathbf x^{k}_{j}, \mathbf u^k_{j} \right) - \bx _{j+1} \|^2_2.
\end{equation}
This minimization is typically performed with some variant of stochastic gradient descent (SGD). As with the linear case, the above methodology can be applied when only measurements $\by _{j}$ are available. MPC using FCNs as a nonlinear prediction model dates back to at least 1995, but has largely fallen out of favor as more advanced neural network architectures have been developed, especially recurrent networks.

Although the models already considered are capable of incorporating time-delays, they do so by augmenting the input dimension of the system model; an $n$ dimensional system modelled with $k$ time-delays will require an input dimension of $nk$. This results in a model that can be computationally inefficient and uninterpretable. A more natural architecture for modelling a dynamical system is that of a recurrent neural network (RNN) which was specifically designed to allow for a sequence of inputs without augmenting the model input dimension. 

RNN architectures function by introducing a latent state, $\bh _{j}$, that evolves according to
\begin{equation}
    \bh _{j+1} = f_{\Theta_r}(\bx _{j}, \bh _{j}), \label{eq:rnn_nocontrol1}
\end{equation}
parameterized by a set of weights and biases $\Theta_r.$ In the context of time-series forecasting, an output operator $f_{\theta_o}$ is learned in conjunction such that $\bh _{j} \mapsto \hat{\bx} _{j}.$ The system 
\begin{equation}
    \bh _{j+1} = f_{\Theta_r} \left( f_{\theta_o}(\bh_j), \bh_{j} \right) \label{eq:rnn_nocontrol2}
\end{equation}
therefore represents an autonomous dynamical system that can be used to generate a forecast of arbitrary length. This interpretation has allowed variants of RNNs to achieve remarkable performance characterizing chaotic dynamical systems, e.g. through Lyapunov exponent calculation and long-term power spectra.

In the control setting, the system of interest is nonautonomous, but the learned RNN can similarly be reduced such that the only exogenous term is the control action, $\bu _{j}$. The dynamical system of the RNN can now be represented by 
\begin{align} \bh _{j+1} &= f_{\Theta_r} \left( \bx _{j}, \bu _{j},\bh _{j}\right)  \label{eq:rnn1}\\
\hat \bx _{j} &= f_{\theta_o} ( \bh _{j} ). \label{eq:rnn2}
\end{align}
An important consequence of such RNN representations of dynamical systems is that the implicit incorporation of time-delays in the model allows for generalization to the case of partial observations $\by _{j}$ without augmenting the size of the network.

The first form of RNN introduced, known as the Elman RNN, assumed forms of $f_{\Theta_r}$ and $f_{\theta_o}$ such that
\begin{align}
\bh _{j+1} &= \sigma_R (W_x \bx _{j} + W_u \bu _{j} + W_h \bh _{j} + b _h) \\
\hat \bx _{j} &= W_{\hat x } \sigma _o ( W_{o} \bh _{j} + b _o) +  b_{\hat x},
\end{align}
where $\Theta_r = \{ W_x, W_u, W_h, b_h\}$ are the trainable recurrent parameters, $\theta_o = \{ W_{\hat x} , W_o, b_o, b_{\hat x}\}$ are the trainable output parameters, and $\sigma_R, \sigma_o$ are the recurrent and output nonlinearities, respectively. Since its introduction, the Elman RNN has largely fallen out of favor because of issues with vanishing/exploding gradients during training. In this work, we consider three variants of RNN inspired by the Elman RNN that help ameliorate these problems: long-short term memory networks (LSTM) and echo state networks (ESN). Both have been used as surrogate models for MPC in the past but have not been extensively compared across a range of benchmark systems. Each considered RNN has the form given in Eqs. \ref{eq:rnn1} and \ref{eq:rnn2}, but differ in their explicit forms of $f_{\Theta_r}$ and $f_{\Theta _o}.$

LSTM networks have long been amongst the most prevalent variants of RNNs, owing to their ability to accurately model systems with long time delays \cite{hochreiter_long_1997}. This ability is a consequence of the gating mechanisms utilized by LSTMs that regulate the flow of information from the input and previous hidden states. For simplicity of presentation, we present the state equations of the LSTM in terms of $\bar \bx _{j}$, which denotes the vector concatenation of $\bx _{j}$ and $\bu _{j}$.
The recurrent mapping of the LSTM is given by
\begin{subequations}
%    \centering
    \begin{align}
    \mathbf{i} _{j} &= \sigma \left( W_{ii} \bar \bx _{j} + b_{ii} + W_{hi} \bh _{j-1} + b_{hi} \right) 
 {\label{eq:input_gate}}\\
 \mathbf{f} _{j} &= \sigma \left( W_{if} \bar \bx _{j} + b_{if} + W_{hf} \bh _{j-1} + b_{hf} \right) 
 {\label{eq:forget_gate}}\\
 \mathbf{g} _{j} &= \tanh \left( W_{ig} \bar \bx _{j} + b_{ig} + W_{hg} \bh _{j-1} + b_{hg} \right) 
 \label{eq:cell_gate}\\
 \mathbf{o} _{j} &= \sigma \left( W_{io} \bar \bx _{j} + b_{io} + W_{ho} \bh _{j-1} + b_{ho} \right) 
 \label{eq:output_gate}\\
 \mathbf{c}_{j} &= \mathbf{f}_{j} \odot \mathbf{c}_{j-1} + \mathbf{i}_{j} \odot \mathbf{g} _{j}
 \label{eq:cell_state}\\
 \bh _{j} &= \mathbf{o}_{j} \odot \tanh \mathbf{c}_{j},
 \label{eq:lstm_hidden_state}
\end{align}
 \label{eq:lstm}
 \end{subequations}%
where $\mathbf i, \mathbf f, \mathbf g$ and $\mathbf o$ denote the input, forget, cell, and output gates, respectively. $\sigma$ represents the sigmoid function and $\odot$ denotes the Hadamard product. Each weight matrix $W$ and bias $b$ are trainable parameters. In addition to propagating the hidden state $\bh$, a so-called cell state $\mathbf c$ is also propagated by the LSTM. From Eq. \ref{eq:cell_state}, we see that $\mathbf c$ only undergoes pointwise modifications, allowing it to act as a ``gradient superhighway.'' For more details on the LSTM architecture, we refer the interested reader to \cite{hochreiter_long_1997}.

In the context of control systems, Eqs. \ref{eq:input_gate}-f represent $f_{\Theta_r}$ from Eq. \ref{eq:rnn1}. We take the output operator $f_{\Theta _o} \left( \mathbf h _{j} \right)$ to be a two-layer feed forward network, akin to that presented in Eq. \ref{eq:fcn}. This is not the only possible choice for $f_{\Theta _o}$, as much of the existing literature takes the output operator to be a single linear layer. However, we found that performance was improved by allowing the output operator to be nonlinear and did not observe notable increases in training time. 

In general, there are two manners in which an LSTM can operate: either stateful or stateless. Stateful RNNs initialize the hidden state once and are trained in a manner such that the evolving hidden state is carried over between batches. In theory, this allows stateful RNNs to learn long-term dependencies. However, stateless RNNs are more frequently used in practice. A stateless RNN will reinitialize the hidden state of the network and use a time-history, usually referred to as lags or time-delays, in order to ``bring the network up to speed'' at the current time-point. We will consider only stateless LSTMs, but comparisons to stateful models is an important area of future work.

One of the most important hyperparameters in the training of an LSTM is the maximal incorporated time-delay, which we denote as $\kappa _1.$ Because we consider stateless LSTMs, $\kappa_1$ sets an upper bound on the length of the ``memory'' of the model. Therefore, $\kappa_1$ determines that $\bh _{j - \kappa} = \mathbf c _{j-\kappa} = \mathbf 0$ at time $j.$ Thus, the hidden state at time $j+1$, $\bh _{j+1},$ is a function of only $\{ (\bx(j-\kappa_1), \bu (j -\kappa_1)), \dots, (\bx _{j}, \bu _{j}) \},$ dictated by $f_{\Theta_r}.$ We seek to minimize the one-step forecast error of the network. As in the case of FCN models, we apply a variant of stochastic gradient descent to identify $\Theta _*$ such that
\begin{equation}
    \Theta_* = \arg \min _\Theta J(\Theta),
\end{equation}
where
\begin{equation}
    J(\Theta) = \sum_j \left\| f_{\theta_o} \left( f_{\Theta_r} \left( \bx _{j}, \bu _{j}, \bh _{j} \right) - \bx _{j+1} \right) \right\|^2_2.
    \label{eq:lstm_train_cost}
\end{equation}
% Once trained, the model is used to forecast multiple time-steps by performing a series of one-step forecasts exactly as in training, but replacing  $\bx _{j}$ with $\hat \bx _{j}$ where necessary. This results in $\kappa_1 T$ forward passes of the recurrent layer. 

In 2014, the gated recurrent unit (GRU) was proposed as an alternative to the LSTM \cite{chung_empirical_2014}. GRUs retain the benefits of the gating mechanisms of LSTMs, but achieve significant reductions in complexity. Rather than propagate two latent variables, $\mathbf c$ and $\mathbf h$, a single latent state $\mathbf h$ is updated. The recurrent mapping is
\begin{subequations}
 %   \centering
\begin{align}
    \mathbf{r} _{j} &= \sigma \left( W_{ir} \bar \bx _{j} + b_{ir} + W_{hr} \bh _{j-1} + b_{hr} \right) 
 \label{eq:cell_gate_gru}\\
    \mathbf z _{j} &= \sigma \left( W_{iz} \bar \bx _{j} + b_{iz} + W_{hz} \bh _{j-1} + b_{hz} \right)
 \label{eq:update_gate}\\
    \mathbf n _{j} &= \tanh [ W_{in} \bar \bx _{j} + b_{in}  + \mathbf r (t) \odot \left( W_{hn} \mathbf h _{j-1} + b_{hn} \right)]
 \label{eq:new_gate}\\
    \bh _{j} &= \left( 1 - \mathbf{z}_{j} \right) \odot \mathbf n _{j} + \mathbf z _{j} \odot \bh _{j-1}.
 \label{eq:gru_hidden_state}
\end{align}
\label{eq:gru}
\end{subequations}
Here $\mathbf r, \; \mathbf z,\;  \text{and}\; \mathbf n$ are called the reset, update, and new gates, respectively. As in the section on LSTMs, each weight matrix $W_{\cdot \cdot}$ and bias $b_{\cdot \cdot}$ are trainable parameters and we have presented the state equations only in terms of an input $\mathbf x$. Similarly, we take the output operator to be a two-layer fully-connected network. % The training and forecasting protocols for all considered GRU forecasters is identical to that of the considered LSTMs.

\section{Methods}
%% intro paragraph
Another class of RNN is the echo state network (ESN) \cite{maass_computational_2004, jaeger_echo_no_date}. While LSTM networks rely on gating mechanisms to control the flow of information and alleviate the problems of vanishing/exploding gradients, ESNs avoid backpropagation altogether. Instead, a random, high-dimensional ``reservoir'' is constructed that is stimulated by a comparably low-dimensional input. The reservoir evolves nonlinearly and it is assumed that the random encodings of this high-dimensional dynamical system can be linearly mapped to the desired low-dimensional output. A trained ESN can then be used as a surrogate model for MPC.

\subsection{ESN Surrogate Model}
Concretely, the recurrent mapping of the ESN is given by
\begin{align}
    \bh _{j+1} = (1-\alpha) \mathbf{h}_{j} +\alpha \tanh \left( A \mathbf{h}_{j} + W_{ih} \bar{\mathbf{x}}_{j} + \sigma _b \mathbf{1} \right), \label{eq:esn_update}
\end{align}
where $\alpha$ is the leak rate, $\mathbf 1$ denotes a vector of ones, and $\sigma_b$ scales the additive bias in the reservoir. Again, $\bar \bx$ denotes the vector concatenation of $\bx$ and $\bu$. $W_{ih}$ and $A$ are randomly initialized and left fixed; there is no direct ``learning'' of the dynamics of the reservoir. Nevertheless, the initialization of $W_{ih}$ and $A$, and the selection of $\alpha$ and $\sigma_b$ play a critical role in the performance of the network. The role of the reservoir matrix $A$ has been especially explored in the literature about ESNs. ESNs rely on a type of generalized synchronization commonly referred to as the echo state property which mandates that given a sufficiently long driving sequence $\bx$, the state of the reservoir is not dependent on $\mathbf h _{0}.$ This desirable property can be achieved through careful initialization of $A.$ In this work, $A$ is selected to be a random, sparse matrix with dimensions $N_r \times N_r$, density $\rho$, and spectral radius $\rho_{sr}.$ In all considered ESNs, $N_r =1000$ and $\rho = 0.02.$ The spectral radius $\rho_{sr}$ is selected by a hyperparameter sweep, although other methods relying on Bayesian optimization have been used to select $\rho_{sr}$ \cite{platt_systematic_2022}. The entries of $W_{ih}$ are drawn from $\mathcal U (-\sigma, \sigma)$ where $\sigma$ is another hyperparameter chosen through a hyperparameter sweep. The parameters $\alpha$ and $\sigma_b$ are similarly chosen.

Despite a larger number of hyperparameters that must be selected, ESNs are typically orders of magnitude faster to train than their LSTM counterparts. This is because only the output operator, $f_{\theta _0}$, is trained and $f_{\theta _0}$ is frequently selected to be linear. Consequently, the training of an ESN amounts to a computationally efficient ridge regression. The reservoir is teacher forced with a sequence of state measurements and control inputs, $\{(\bx _{0}, \bu _{0}), \dots, (\bx _{N}, \bu_{N}\}$ to generate a corresponding sequence $\{ \bh (1), \dots, \bh (N+1) \}$ where $\bh _{0} = \mathbf{0}$. Here, the echo state property ensures that our selection of $\bh _{0}$ will be ``forgotten'' after a sufficiently long sequence of teacher forcing. In general, the minimum length of this sequence depends on the problem at hand, but in all the examples we consider, $N_{spin} = 200$ time-steps appear sufficient. This is substantially fewer than suggested in some existing literature, perhaps as a consequence of an external driving force as opposed to autonomous dynamics. Larger $N_{spin}$ effectively limits the amount of data on which the output operator can be trained. $f_{\Theta _o}$ takes the form
\begin{equation}
    f_{\theta _o} (\bh _{j}) = W_{o} \bh _{j}
\end{equation}
and we determine $W_{o}$ by performing a Ridge regression on
\begin{equation}
    W_{o} \begin{bmatrix}
        | & & |\\
        \bh _{N_{spin}} & \cdots & \bh _{N+1}\\
        | & & |
    \end{bmatrix} = \begin{bmatrix}
        | & & |\\
        \bx _{N_{spin}} & \cdots & \bx _{N+1}\\
        | & & |
    \end{bmatrix}.
\end{equation}
The Tikhonov regularization parameter $\beta$ is another hyperparameter that is tuned using grid search. Some previous works have relaxed the assumption that the output operator be linear and take $f_{\theta _o}$ to include quadratic nonlinearities in some entries of $\bh _{j}$. In this form of an ESN, a ridge regression is still used for training. However, recent work has shown that this inclusion is unnecessary when $\sigma_b$ is allowed to be non-zero, as is the case here. 

This training protocol performs a minimization of the error on a one-step forecast in much the same way as Eq. \ref{eq:lstm_train_cost} does for the LSTM. A multi-step forecast is achieved by iterating Eq. \ref{eq:esn_update} as
\begin{equation}
    \bh _{j+1} = f_{\Theta_r} ( \hat \bx _{j}, \bu _{j},\bh _{j}).
\end{equation}
An advantage of this form is that to perform a $T$-step forecast, $T$ forward passes of the network are required; no reinitialization of the reservoir state $\bh$ is ever needed. In this manner, an ESN maintains the benefits of a stateful RNN without the additional training difficulties stateful models introduce in RNN architectures requiring backpropagation and gradient descent.

\subsection{Training Protocols}
The training of surrogate models for each example system was performed in the same manner. For any example system, let
\begin{equation}
    \mathbf{X} = \begin{bmatrix}
        | & & | \\
        \bx_0 & \cdots & \bx_N \\
        | & & |
    \end{bmatrix}
\end{equation}
denote the training data matrix. The data matrix was then corrupted column-wise with Gaussian white noise with covariance 
\begin{equation}
    \Sigma ^2= \begin{bmatrix}
        0.001 \sigma_{tr,0}^2 & 0 & \cdots & 0 \\
        0 & 0.001\sigma_{tr,1}^2 & \cdots & 0 \\
        \vdots & \vdots & \ddots & 0 \\
        0 & 0 & \cdots & 0.001\sigma_{tr,N_{x}} ^2
    \end{bmatrix}
\end{equation}
where $\sigma _{tr}$ is a vector containing the componentwise empirical standard deviation of the training data. The corruption of training data in this manner acts as a sort of regularization and has been shown to be beneficial in the downstream forecasting performance of surrogate models. The data is then min-max scaled row-wise. 

For each system, a validation dataset was also generated, which will be discussed at greater length in the following section. An extensive hyperparameter sweep for each candidate class of surrogate model was then performed. Given a set of hyperparameters for a candidate model, we train the model on the set of training data and evaluate its 50 step RMSE forecast performance on the validation data. 50 step forecast performance was chosen because longer forecasts render the online optimization in MPC computationally prohibitive. In the case of the ESN and DMDc models, one of the hyperparameters over which the sweep is performed is $\beta$, the Tikhonov regularization parameter used for the ridge regression in training. Because this parameter on its own can reduce overfitting, 100\% of the training data was used in training these models.

On the other hand, the models trained via backpropagation and the Adam optimizer do not have a hyperparameter that directly controls overfitting to the data. As a result, during training 20\% of the training data is withheld to serve as a first validation set. An early stopping criteria is introduced based on performance in this validation set to prevent overfitting.

From the many models trained in the hyperparameter search, we select the best performing set of hyperparameters for subsequent forecast and control performance demonstrations. However, we do not use the specific model found during the hyperparameter sweep. This is because the random initializations of neural networks can have a significant impact on subsequent performance, potentially causing the continued use of the model from the hyperparameter sweep to yield better results than can be expected in most cases. We note that although we consider significantly more hyperparameter combinations for the ESN compared to the LSTM, the total time required to perform the sweep is still much larger for the LSTM. This is a consequence and benefit of the rapid training enabled by reservoir computing approaches.

\subsection{MPC Formulation}
In all cases, we define the MPC cost function as 
\begin{align}
    J_{tot} (U,t) = J(U,t) + \alpha_1 \| \max (U, b_1) - b_1 \| ^2 \\
    + \alpha_1 \| \min (U, b_2) - b_2 \| ^2  \notag,
    \label{eq:tot_mpc}
\end{align}
where $J(U,t)$ is defined in Eq. \ref{eq:mpc_cost}. The additional terms represent a type of soft control barrier function. The constant matrices $b_1$ and $b_2$ are chosen with entries to be 0.95 and 0.05, respectively, for all datasets and the max/min functions operate elementwise \cite{ames_control_2019}. The same constants can be used for all benchmark systems because the models perform forecasts in  min-max scaled data. Similarly, $\alpha_1 = 100.$ Because the introduced CBF does not impose hard constraints, any entries of $U$ found by the numerical solver are clamped to the range [0,1]. The introduction of the soft constraints encourages the solver to identify control actions between 0.05 and 0.95, while the hard constraints ensure that the maximum/minimum control values do not exceed those found in the training data. The forecast horizon and control horizons are selected to be $T = M = 50$, while the number of control steps used from a single optimization is taken to be $M_c = 20.$ These selections provide a more stringent test of the learned surrogate models, as even very poor models can be effective with MPC when $T, M,$ and $M_c$ are small. Finally, the weighting matrices for $J(U,t)$ are selected as $Q_1 = 100 \mathbb I$, $Q_2 = 1 \mathbb I$, and $Q_3 = 20 \mathbb I$. $Q_1$ determines the cost of deviation from the desired reference trajectory, while $Q_2$ and $Q_3$ penalize the magnitude and the magnitude of the derivative of control actions. Following existing work using RNNs for MPC, we use an L-BFGS algorithm to perform the required online minimizations of Eq. \ref{eq:tot_mpc} \cite{bieker_deep_2020}.

\section{Forecasting Results}
\begin{figure*}[!htb]
    \centering
    \begin{overpic}[width=0.77\textwidth]{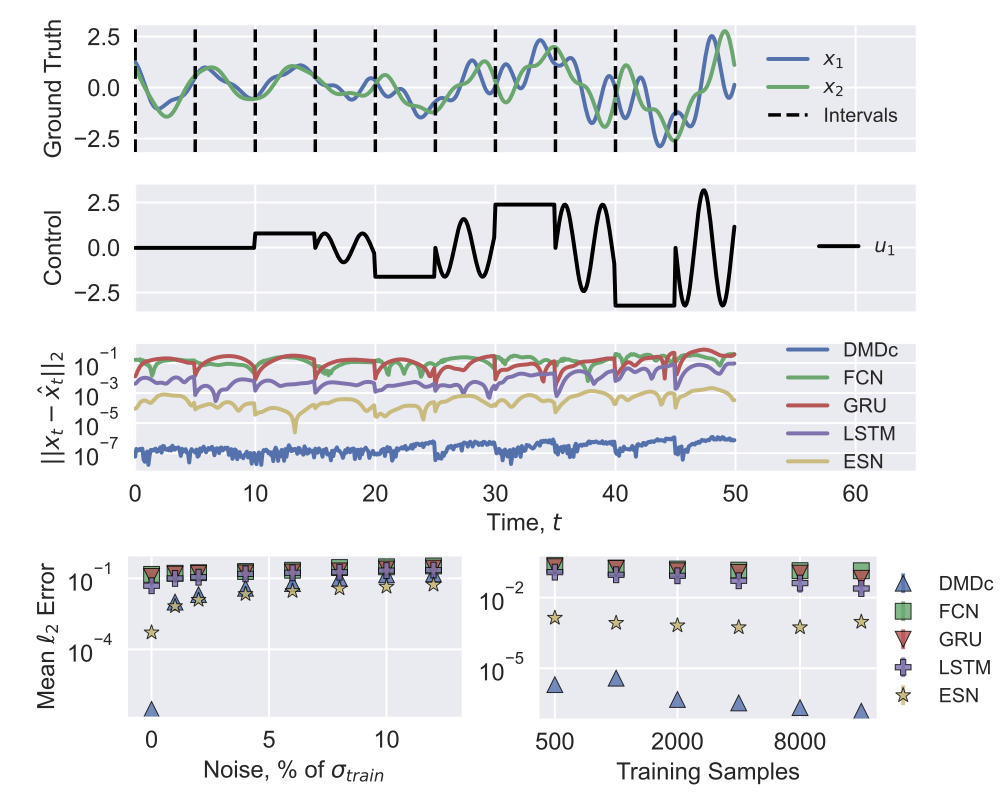} 
    \put(1, 79){\large \textbf{A}}
    \put(1,25){\large \textbf{B}}
    \put(50, 25){\large \textbf{C}}
    \end{overpic}
    \caption{Surrogate model forecasting results for the spring mass system. Panel A shows the time-series of the validation data (top) and the corresponding control inputs (middle). The time-series of $\mathbf{x}$ has been scaled to improve readability across examples. The bottom of Panel A shows the $\ell_2$ deviation of each surrogate model's forecast from the ground truth as a function of time. All forecasts are reinitialized every 50 time-steps, denoted by the dashed black line in the top plot. Panel B reports the mean $\ell_2$ forecast error obtained across 32 trained surrogate models of each type for varying levels of additive Gaussian white noise. Panel C reports analogous results for the case of zero noise, but a varying number of training samples.}
    \label{fig:spring_mass_forecasting}
\end{figure*}
\begin{figure*}[!htb]
    \centering
    \begin{overpic}[width=0.77\textwidth]{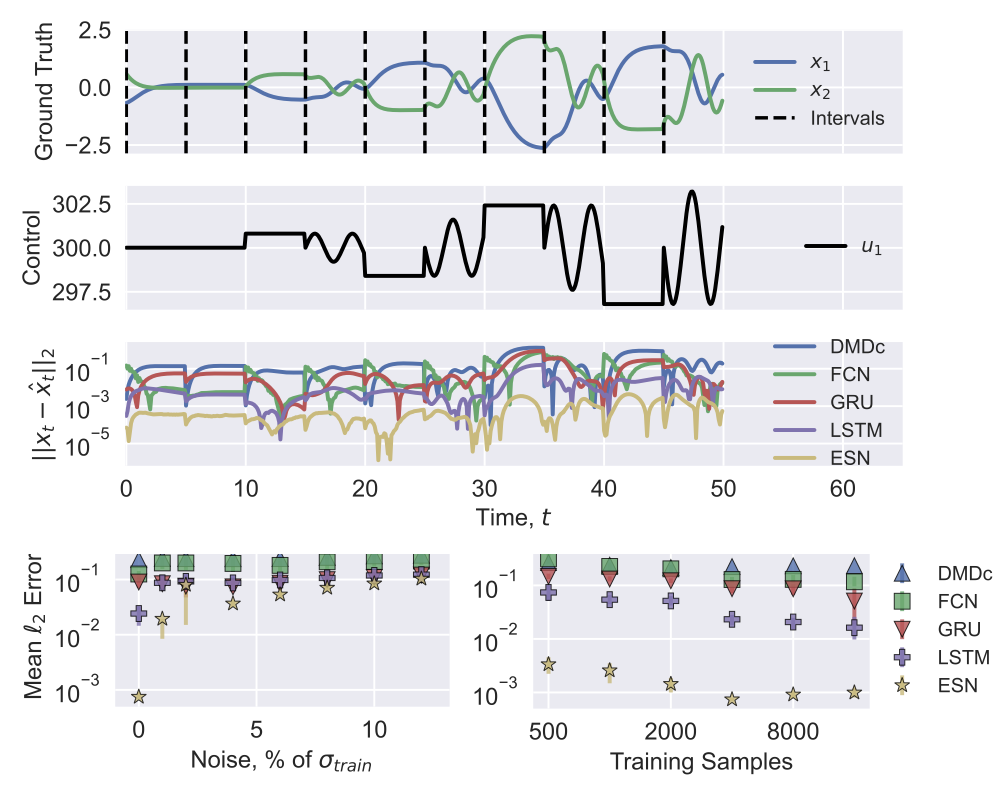} 
    \put(1, 79){\large \textbf{A}}
    \put(1,25){\large \textbf{B}}
    \put(50, 25){\large \textbf{C}}
    \end{overpic}
    \caption{Surrogate model forecasting results for the stirred tank system. All panels are as labelled in Fig. \ref{fig:spring_mass_forecasting}. Panel A shows the time-series of the validation data (top) and the corresponding control inputs (middle). The time-series of $\mathbf{x}$ has been scaled to improve readability across examples. The bottom of Panel A shows the $\ell_2$ deviation of each surrogate model's forecast from the ground truth as a function of time. All forecasts are reinitialized every 50 time-steps, denoted by the dashed black line in the top plot. Panel B reports the mean $\ell_2$ forecast error obtained across 32 trained surrogate models of each type for varying levels of additive Gaussian white noise. Panel C reports analogous results for the case of zero noise, but a varying number of training samples.}
    \label{fig:stirred_tank_forecasting}
\end{figure*}
\begin{figure*}[!htb]
    \centering
    \begin{overpic}[width=0.77\textwidth]{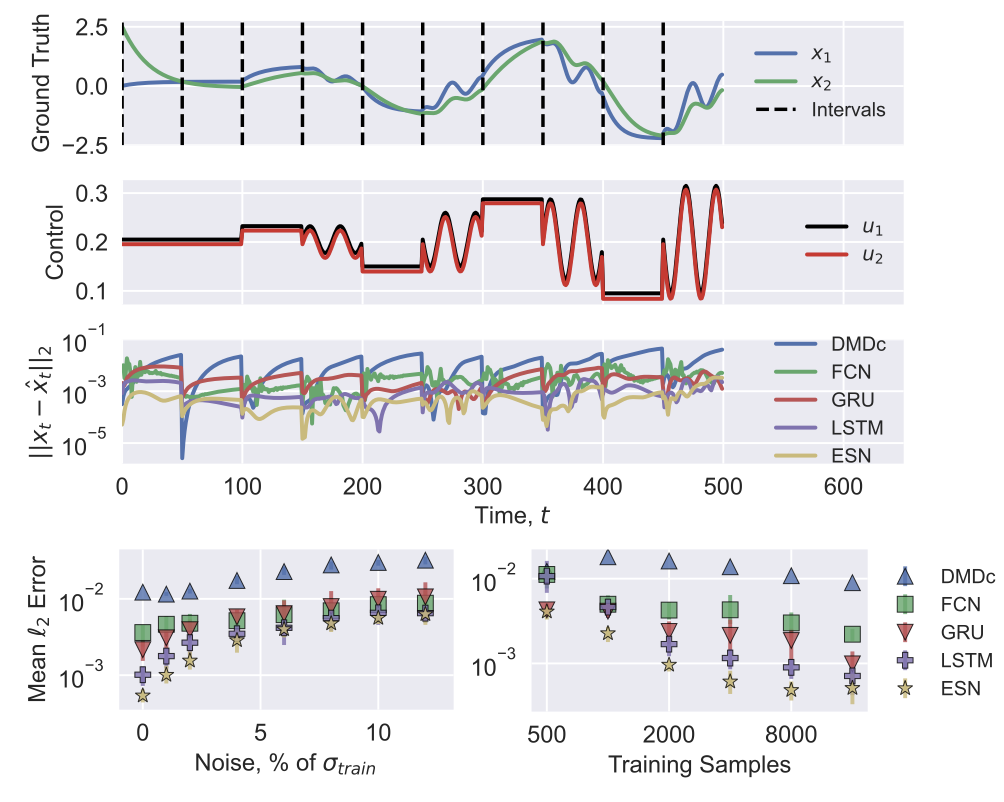} 
    \put(1, 79){\large \textbf{A}}
    \put(1,25){\large \textbf{B}}
    \put(50, 25){\large \textbf{C}}
    \end{overpic}
    \caption{Surrogate model forecasting results for the two tank reservoir system. All panels are as labelled in Fig. \ref{fig:spring_mass_forecasting}. Panel A shows the time-series of the validation data (top) and the corresponding control inputs (middle). The time-series of $\mathbf{x}$ has been scaled to improve readability across examples. The bottom of Panel A shows the $\ell_2$ deviation of each surrogate model's forecast from the ground truth as a function of time. All forecasts are reinitialized every 50 time-steps, denoted by the dashed black line in the top plot. Panel B reports the mean $\ell_2$ forecast error obtained across 32 trained surrogate models of each type for varying levels of additive Gaussian white noise. Panel C reports analogous results for the case of zero noise, but a varying number of training samples.}
    \label{fig:two_tank_forecasting}
\end{figure*}
\begin{figure*}[!htb]
    \centering
    \begin{overpic}[width=0.77\textwidth]{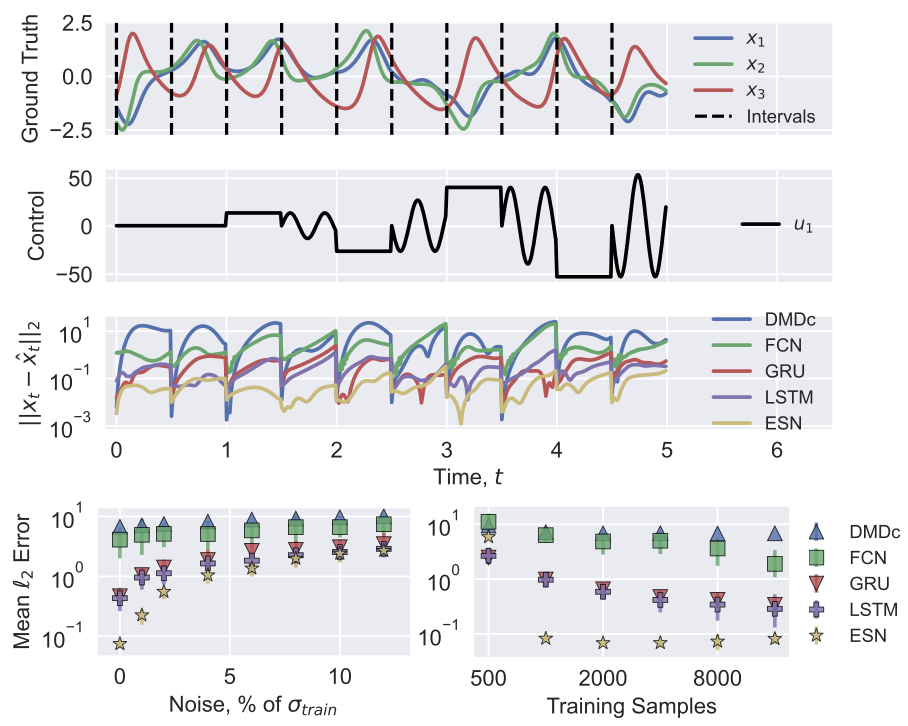} 
    \put(1, 79){\large \textbf{A}}
    \put(1,24){\large \textbf{B}}
    \put(48, 24){\large \textbf{C}}
    \end{overpic}
    \caption{Surrogate model forecasting results for the Lorenz system with control. All panels are as labelled in Fig. \ref{fig:spring_mass_forecasting}. Panel A shows the time-series of the validation data (top) and the corresponding control inputs (middle). The time-series of $\mathbf{x}$ has been scaled to improve readability across examples. The bottom of Panel A shows the $\ell_2$ deviation of each surrogate model's forecast from the ground truth as a function of time. All forecasts are reinitialized every 50 time-steps, denoted by the dashed black line in the top plot. Panel B reports the mean $\ell_2$ forecast error obtained across 32 trained surrogate models of each type for varying levels of additive Gaussian white noise. Panel C reports analogous results for the case of zero noise, but a varying number of training samples.}
    \label{fig:lorenz_forecasting}
\end{figure*}

\begin{figure*}[!htb]
    \centering
    \begin{overpic}[width=0.77\textwidth]{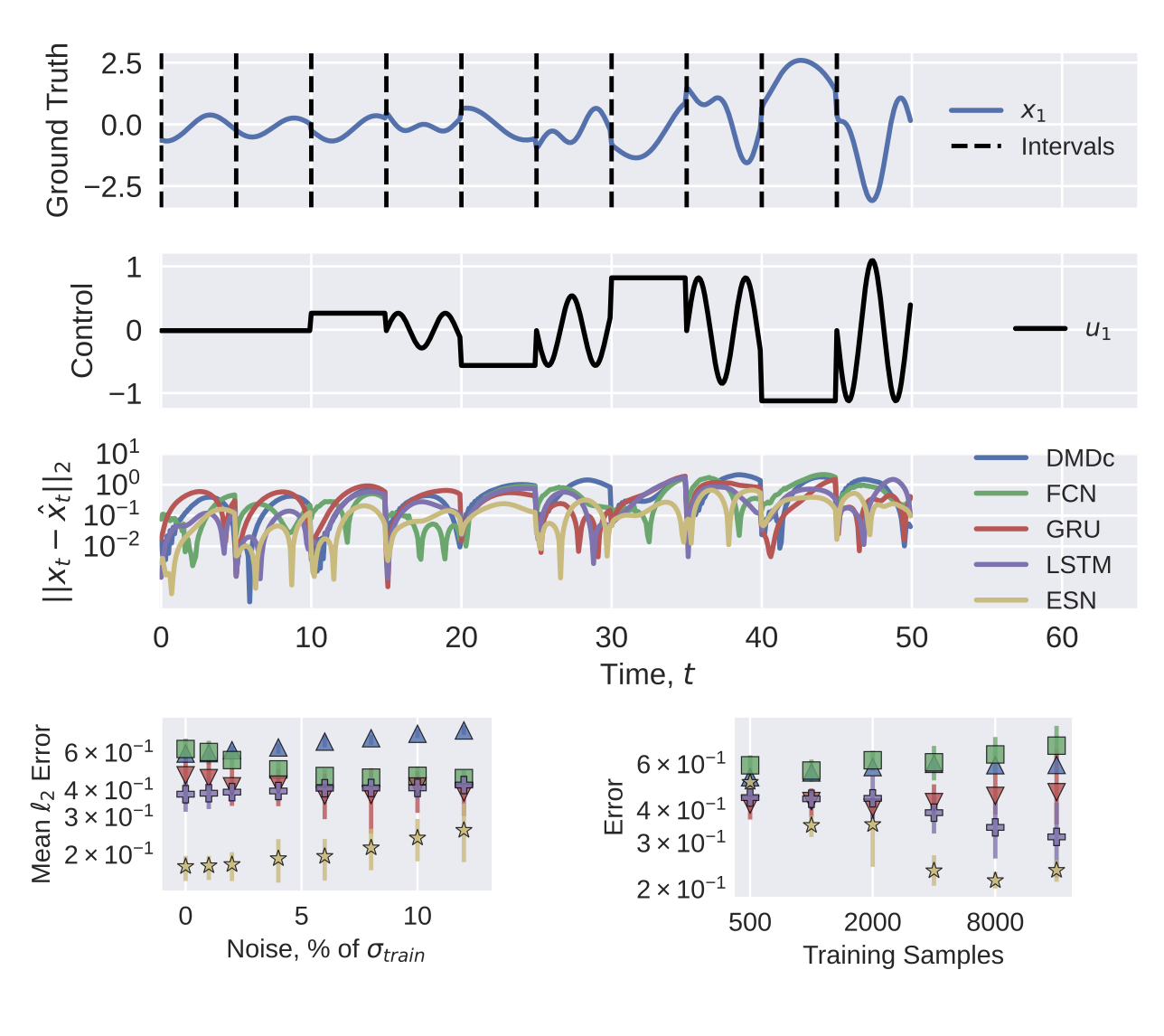} 
    \put(1, 90){\large \textbf{A}}
    \put(1,28){\large \textbf{B}}
    \put(60, 28){\large \textbf{C}}
    \end{overpic}
    \caption{Surrogate model forecasting results for flow past a cylinder. Panel A shows the time-series of the validation data (top) and the corresponding control inputs (middle). The time-series of $\mathbf{x} = C_{\ell}$ has been scaled to improve readability across examples. The bottom of Panel A shows the $\ell_2$ deviation of each surrogate model's forecast from the ground truth as a function of time. All forecasts are reinitialized every 50 time-steps, denoted by the dashed black line in the top plot. Panel B reports the mean $\ell_2$ forecast error obtained across 32 trained surrogate models of each type for varying levels of additive Gaussian white noise. Panel C reports analogous results for the case of zero noise, but a varying number of training samples.}
    \label{fig:cyl_forecasting}
\end{figure*}
\begin{figure}[!htb]
    \centering
    \includegraphics[width=\linewidth]{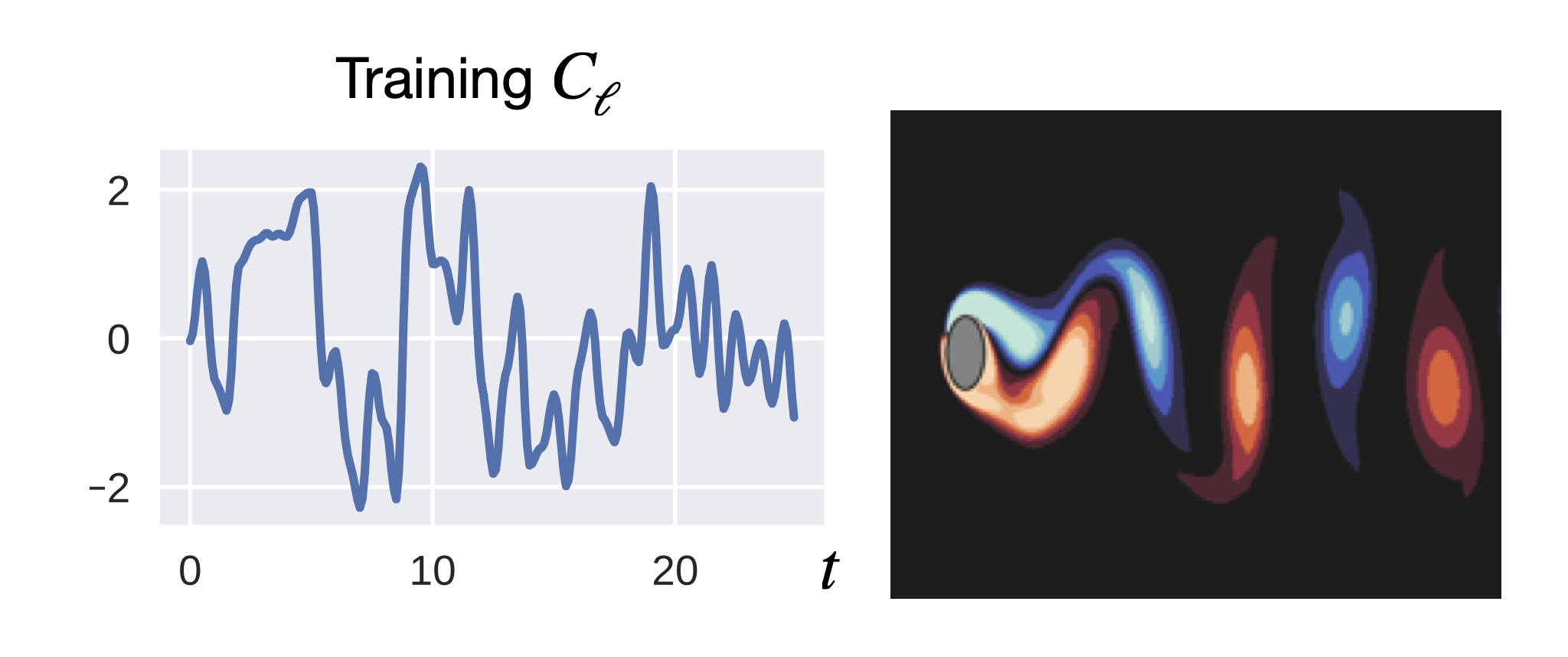}
    \caption{Flow past a cylinder visualization. (Left) shows the flow field of a training snapshot and (right) illustrates a selection of the generated training data.}
    \label{fig:cyl_viz}
\end{figure}
A ubiquitous issue in the training of surrogate models for control systems is that the structure of the control signal used for training does not mimic that of effective control. While one may find that a trained surrogate model performs well on data that resembles that seen in training, the model may prove ineffective for different control signals. For this reason, for each example system we construct a validation dataset with control inputs that switch between constants and smoothly varying sinusoids. Such control signals are not seen in training, where the control resembles a series of rapidly varying step inputs. Although the validation control signal also does not mimic effective control, it provides a first test of a model's ability to function under previously unseen control inputs. 

In a departure from existing work on surrogate models for autonomous systems, we emphasize the accuracy of short-term, 50-step forecasts, as this ability is most relevant to subsequent MPC. The selection of 50-step forecasts is somewhat arbitrary. In this work, it was selected because it represents an approximate upper bound on the forecast horizon for which MPC is implementable with the discretizations of the majority of considered models.

\subsection{Linear, spring mass system}

The first example control system is inspired by a simple, coupled spring-mass system. The continuous SIMO system is represented by the following system state-space representation:
\begin{subequations}
\begin{align}
    \dot \bx &= \begin{bmatrix}
        0 & 1 & 0 & 0\\
        \frac{-2 k}{m} & \frac{-b}{m} & \frac{k}{m} & 0 \\
        0 & 0 & 0 & 1\\
        \frac{k}{m} & 0 & \frac{-k}{m} & \frac{-b}{m}
    \end{bmatrix} \bx + \begin{bmatrix}
        0 \\
        \frac{1}{m}\\
        0 \\
        0
    \end{bmatrix} \bu \\
    \by &= \begin{bmatrix}
        1 & 0 & 1 & 0
    \end{bmatrix} \bx,
\end{align}    
\end{subequations}

where $k$, $m$, and $b$ are constants. The control discretization is taken to be $\Delta t = 0.1$s. Training data is generated by simulating the system under a pseudo-random control signal for 5000 seconds, corresponding to 50,000 temporal snapshots. The control signal was generated a priori by drawing samples from $\mathcal U (-3,3)$ at intervals of 0.5 seconds and repeating each drawn value for 5 time-steps. To obtain a more physical control signal, the signal was then convolved with a uniform filter of width 2. 

The validation control signal and the performance of a particular instance of each candidate model for the linear, spring mass system is shown in Panel A of Fig. \ref{fig:spring_mass_forecasting}. In the case of zero noise, as in Panel A and C, DMDc models achieve the best performance because the underlying system is linear and observable. Thus, the DMDc models inherently encode the best possible representation of the dynamics. ESNs provide the second best approximation of the dynamics with no noise, and outperform all competing methods in the presence of all considered, non-zero noise levels. All candidate models achieve similar performance when the number of training samples is varied between 500 and 32,000. The hyperparameters of each model are given in Table \ref{tab:spring_mass_forecasting}.

\begin{table}[!htb]
    \centering
    \begin{tabular}{|c|c|}
        \hline
        DMDc & $N_l = 20$, $\beta = 10^{-6}$ \\
        \hline
        FCN & $N_l = 15,$, $p_d = 0.0,$ $N_w = 50$, $\alpha_{lr} = 10^{-3}$ \\
        \hline
        GRU & $N_r = 64$, $p_d = 0.0$, $N_l = 20$, $\alpha_{lr} = 10^{-3}$\\
        \hline
        LSTM & $N_r = 128$, $p_d = 0.0$, $N_l = 30$, $\alpha_{lr} = 10^{-3}$ \\
        \hline
        ESN & $\rho_{sr} = 0.8$, $\sigma = 0.01$, $\sigma_b = 0.66$, $\alpha = 0.2$, $\beta = 10^{-7}$\\
        \hline
    \end{tabular}
    \caption{Selected hyperparameters for the spring mass system.}
    \label{tab:spring_mass_forecasting}
\end{table}

\subsection{Continuous stirred tank reactor}

The second example system we consider is also a SIMO system, but now includes non-linearities. The control aim of the system is to maintain the temperature of a chemical reaction by modulating the jacket temperature of the reactor. We let $x_1$ denote the concentration of a chemical $A_{chem}$ and $x_2$ be the reactor temperature. $u$ denotes the jacket temperature. The example system was proposed by AP Monitor and we rely on the Python package neuromancer for simulation. The control discretization is set as $\Delta t = 0.1$s and the system was simulated for 5,000s. The control signal used for the generation of training data was constructed by drawing samples from $\mathcal U (297, 303)$ to mimic the operating parameters of the cooling jacket. Samples were drawn every 0.5s and held constant until the next sample was drawn. Again, we apply a smoothing uniform filter of width 2.

Fig. \ref{fig:stirred_tank_forecasting} reports results for the continuous stirred tank reactor system analogous to Fig. \ref{fig:spring_mass_forecasting} for the spring mass system. The hyperparameters of each model are given in Table \ref{tab:stirred_tank_forecasting}. Because of the inherently nonlinear dynamics, DMDc models no longer are appropriate for modelling the system and yield the worst performance. ESNs achieve the best forecasting performance across all considered noise levels and number of training samples. As before, the gap in performance is largest in the case of zero noise and diminishes as noise is increased. LSTMs consistently provide the second best surrogate model. All models exhibit moderate reductions in mean $\ell_2$ forecasting error as the number of training samples is increased. 

\begin{table}[!h]
    \centering
    \begin{tabular}{|c|c|}
        \hline
        DMDc & $N_l = 20$, $\beta = 10^{-4}$ \\
        \hline
        FCN & $N_l = 15,$, $p_d = 0.0,$ $N_w = 10$, $\alpha_{lr} = 10^{-4}$ \\
        \hline
        GRU & $N_r = 128$, $p_d = 0.02$, $N_l = 30$, $\alpha_{lr} = 10^{-4}$\\
        \hline
        LSTM & $N_r = 128$, $p_d = 0.0$, $N_l = 20$, $\alpha_{lr} = 10^{-4}$ \\
        \hline
        ESN & $\rho_{sr} = 0.4$, $\sigma = 0.25$, $\sigma_b = 1.33$, $\alpha = 0.4$, $\beta = 10^{-6}$\\
        \hline
    \end{tabular}
    \caption{Selected hyperparameters for the stirred tank system.}
    \label{tab:stirred_tank_forecasting}
\end{table}

\subsection{Two-tank reservoir}
We now turn to a nonlinear, MIMO system consisting of coupled water tanks. Water can be pumped into either tank independently, $u_1$ and $u_2$. Tank 1 drains into Tank 2, and Tank 2 drains into a large reservoir. The water level in each tank are taken as the state variables $x_1, x_2$ and control aims to achieve set-points in each tank. As in the case of the stirred tank reactor, the system was proposed by AP Monitor and we rely on the package neuromancer for simulation. The dynamics occur on longer time-scales than in the previous examples, leading us to select $\Delta t = 1$s as the control discretization. In the generation of training data, both $u_1$ and $u_2$ were drawn independently from $\mathcal U (0, 0.4)$ at intervals of 50 seconds for 50,000 seconds. A smoothing uniform filter of width 50 was then applied entrywise.

Fig. \ref{fig:two_tank_forecasting} and Table \ref{tab:two_tank_forecasting} report the forecasting performance and selected hyperparameters, respectively, for each considered model in the two-tank reservoir system. ESNs provide the best approximation of the dynamics across all noise-levels and number of training samples, although the disparity between ESNs and their LSTM counterparts is greatly reduced in this system. All models display greater reductions in error as the number of training samples is increased than in the two previous systems. This is likely a result of the MIMO nature of the two-tank system requiring more data to fully explore the state space of the system. 
\begin{table}[!h]
    \centering
    \begin{tabular}{|c|c|}
        \hline
        DMDc & $N_l = 20$, $\beta = 10^{-7}$ \\
        \hline
        FCN & $N_l = 20,$, $p_d = 0.0,$ $N_w = 50$, $\alpha_{lr} = 10^{-3}$ \\
        \hline
        GRU & $N_r = 128$, $p_d = 0.0$, $N_l = 5$, $\alpha_{lr} = 10^{-4}$\\
        \hline
        LSTM & $N_r = 16$, $p_d = 0.0$, $N_l = 20$, $\alpha_{lr} = 10^{-4}$ \\
        \hline
        ESN & $\rho_{sr} = 0.8$, $\sigma = 0.5$, $\sigma_b = 1.66$, $\alpha = 0.2$, $\beta = 10^{-5}$\\
        \hline
    \end{tabular}
    \caption{Selected hyperparameters for the two-tank reservoir system.}
    \label{tab:two_tank_forecasting}
\end{table}

\subsection{Lorenz system with control}
Proposed in 1963 by Edward Lorenz, the Lorenz system is frequently used as an example of chaotic dynamics. Indeed, many existing works using RNNs for the forecasting of autonomous dynamical systems demonstrate algorithmic performance on the Lorenz system. Here, we augment the classic Lorenz system with a control term $u$
\begin{subequations}
\begin{align}
    \dot{x}_1 &= \sigma (x_2 - x_1) + u\\
    \dot{x}_2 &= x_1 (\rho - x_3) - x_2\\
    \dot{x}_3 &= x_1 x_2 - \beta x_3,
\end{align}
\end{subequations}
where $\sigma = 10$, $\rho = 28$, and $\beta = 8/3.$ The control task will be to successively steer between the three fixed points of the autonomous Lorenz system, and the considered control discretization is $\Delta t = 0.01$s. Following the work of Kaiser et al. \cite{kaiser_sparse_2018}, we generate training data by sampling $u$ from $\mathcal U(-50, 50)$ every 0.05 seconds for 500 seconds and smoothing the control signal with a uniform filter of width 2. 

Fig. \ref{fig:lorenz_forecasting} and Table \ref{tab:lorenz_forecasting} report the forecasting performance and selected hyperparameters, respectively, for each considered model in Lorenz system with control. In contrast to the previous systems, the Lorenz system exhibits chaotic behavior rendering forecasts significantly more challenging, as exhibited by the higher mean $\ell_2$ error for all models. Consequently, all models fail to achieve accurate forecasts when only 500 training samples are available. However, with 1,000 training samples, ESNs outperform all competing models with as many as 32,000 training samples. Beyond low-levels of noise, all models fail to perform accurate forecasts, but ESNs still provide the best approximation. 
\begin{table}[!htb]
    \centering
    \begin{tabular}{|c|c|}
        \hline
        DMDc & $N_l = 20$, $\beta = 10^{-6}$ \\
        \hline
        FCN & $N_l = 15,$ $p_d = 0.0,$ $N_w = 75$, $\alpha_{lr} = 10^{-4}$ \\
        \hline
        GRU & $N_r = 32$, $p_d = 0.0$, $N_l = 10$, $\alpha_{lr} = 10^{-4}$\\
        \hline
        LSTM & $N_r = 128$, $p_d = 0.0$, $N_l = 30$, $\alpha_{lr} = 10^{-4}$ \\
        \hline
        ESN & $\rho_{sr} = 0.4$, $\sigma = 0.1$, $\sigma_b = 1.33$, $\alpha = 0.2$, $\beta = 10^{-7}$\\
        \hline
    \end{tabular}
    \caption{Selected hyperparameters for the Lorenz system with control.}
    \label{tab:lorenz_forecasting}
\end{table}

\subsection{Flow past a cylinder}

The final system under consideration is a fluid flow past a cylinder. The governing Navier-Stokes equations are given by
\begin{subequations}
\begin{align}
    \frac{\partial u}{\partial t}+u \cdot (\nabla u) &= -\nabla p + \frac{1}{\text{Re}} \Delta u \\
    \nabla \cdot u &= 0,
\end{align}
\end{subequations}
where $u,$ $t,$ and $p$ are nondimensional velocity, time, and pressure, respectively, and \textit{Re} is the Reynolds number which is set to 100. A sample visualization of the flow is given in Fig. \ref{fig:cyl_viz} A. 

Control is accomplished by rotating the cylinder, and we assume a no-slip condition at the interface of the cylinder and the fluid. The control objective is to control the lift generated by the cylinder, and all surrogate models are constructed assuming only access to lift measurements. Training data is generated by simulating the system under a pseudo-random control sequence with values drawn from $\mathcal U (-\pi/2, \pi/2)$ for 5,000 seconds with a control discretization of $\Delta t=0.1$s. Control values are selected every 0.5s and the resulting signal is smoothed with a uniform filter of width 2. Fig. \ref{fig:cyl_viz} B shows the evolution of the coefficient of lift, $C_{\ell}$, for a 25 second period of the training data.

Fig. \ref{fig:cyl_forecasting} and Table \ref{tab:cyl_forecasting} display the forecasting results and selected hyperparameters for each considered class of surrogate model. Once again, ESNs consistently outperform all competing methods, albeit by a smaller margin in this case.

\begin{table}[!htb]
    \centering
    \begin{tabular}{|c|c|}
        \hline
        DMDc & $N_l = 20$, $\beta = 0.0$ \\
        \hline
        FCN & $N_l = 20,$ $p_d = 0.0,$ $N_w = 10$, $\alpha_{lr} = 10^{-2}$ \\
        \hline
        GRU & $N_r = 16$, $p_d = 0.0$, $N_l = 30$, $\alpha_{lr} = 10^{-2}$\\
        \hline
        LSTM & $N_r = 16$, $p_d = 0.0$, $N_l = 20$, $\alpha_{lr} = 10^{-3}$ \\
        \hline
        ESN & $\rho_{sr} = 0.4$, $\sigma = 0.5$, $\sigma_b = 0.33$, $\alpha = 0.2$, $\beta = 10^{-7}$\\
        \hline
    \end{tabular}
    \caption{Selected hyperparameters for the flow past a cylinder.}
    \label{tab:cyl_forecasting}
\end{table}

\section{ESN Control Results}
\begin{figure*}[!htb]
    \centering
    \begin{overpic}[width=0.77\textwidth]{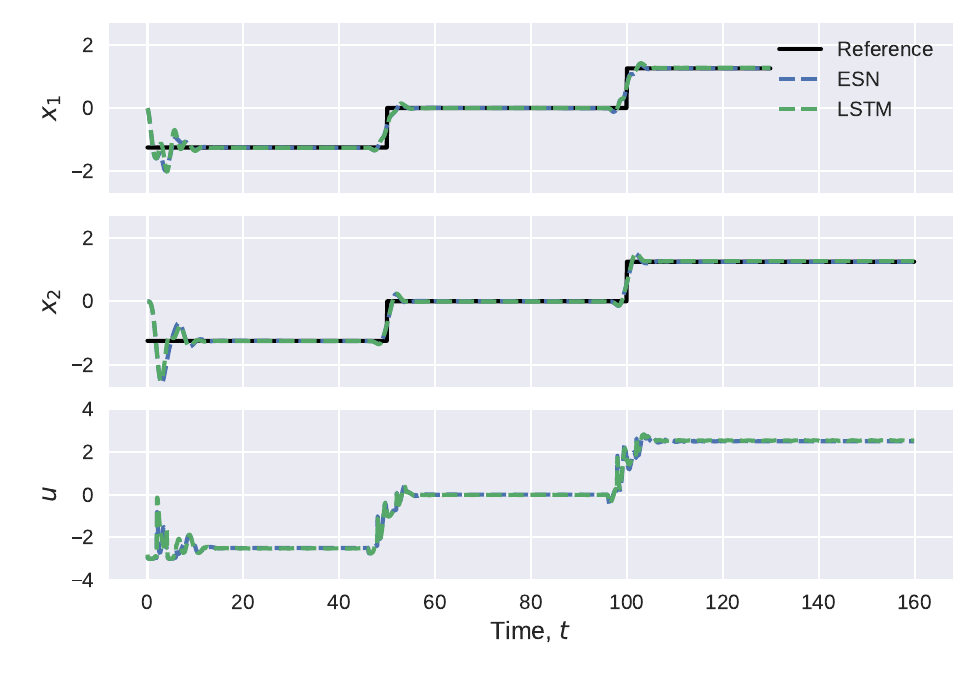} 
    \end{overpic}
    \caption{Control performance of LSTM and ESN based MPC in the spring mass system. The aim of each controller is to successively steer an arbitrary initial condition to three set point values.}
    \label{fig:springmass_control}
\end{figure*}
\begin{figure*}[!htb]
    \centering
    \begin{overpic}[width=0.77\textwidth]{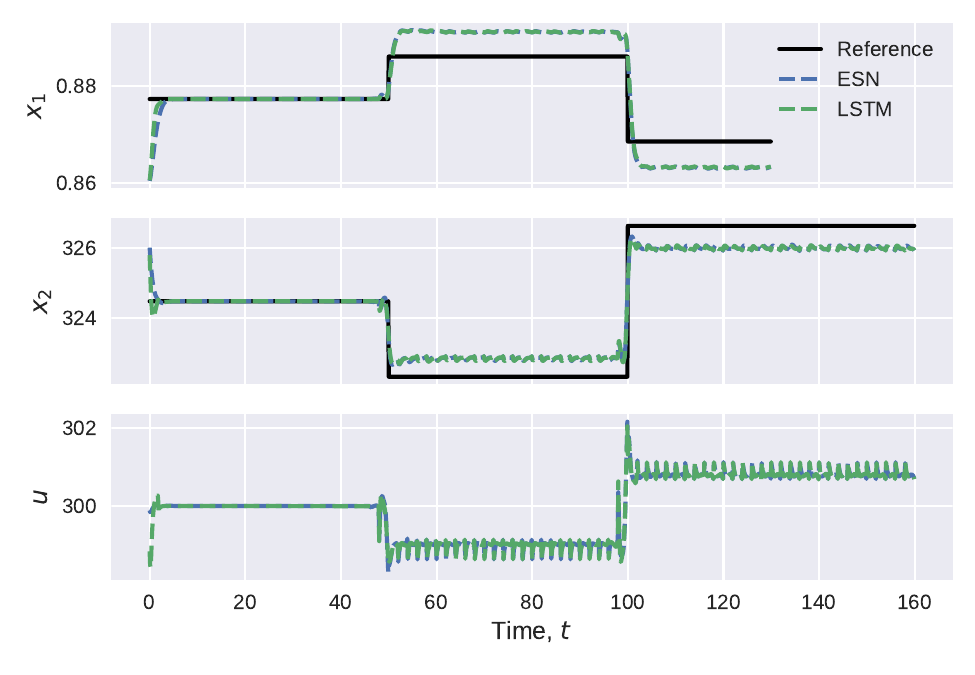} 
    \end{overpic}
    \caption{Control performance of LSTM and ESN based MPC in the stirred tank system. The aim of each controller is to successively steer an arbitrary initial condition to three set point values. In this case, the set point values have been selected such that the first set point is feasible, while the second two are not. This provides a test of the methods' robustness to infeasible reference trajectories.}
    \label{fig:stirred_tank_control}
\end{figure*}
\begin{figure*}[!htb]
    \centering
    \begin{overpic}[width=0.77\textwidth]{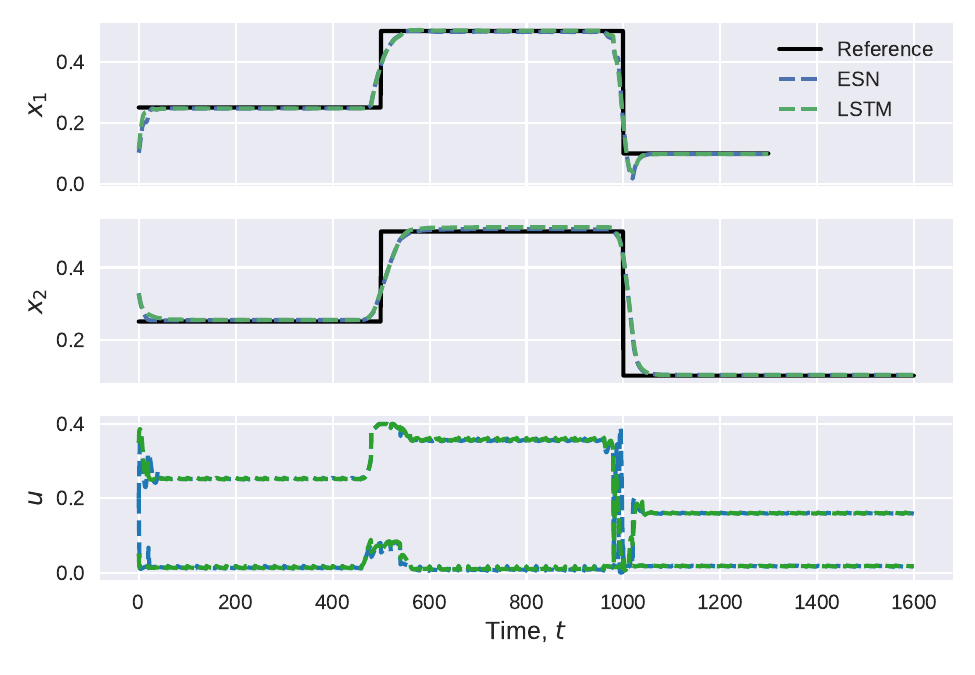} 
    \end{overpic}
    \caption{Control performance of LSTM and ESN based MPC in the two tank reservoir system. The aim of each controller is to successively steer an arbitrary initial condition to three set point values. }
    \label{fig:two_tank_control}
\end{figure*}
\begin{figure*}[!htb]
    \centering
    \begin{overpic}[width=0.77\textwidth]{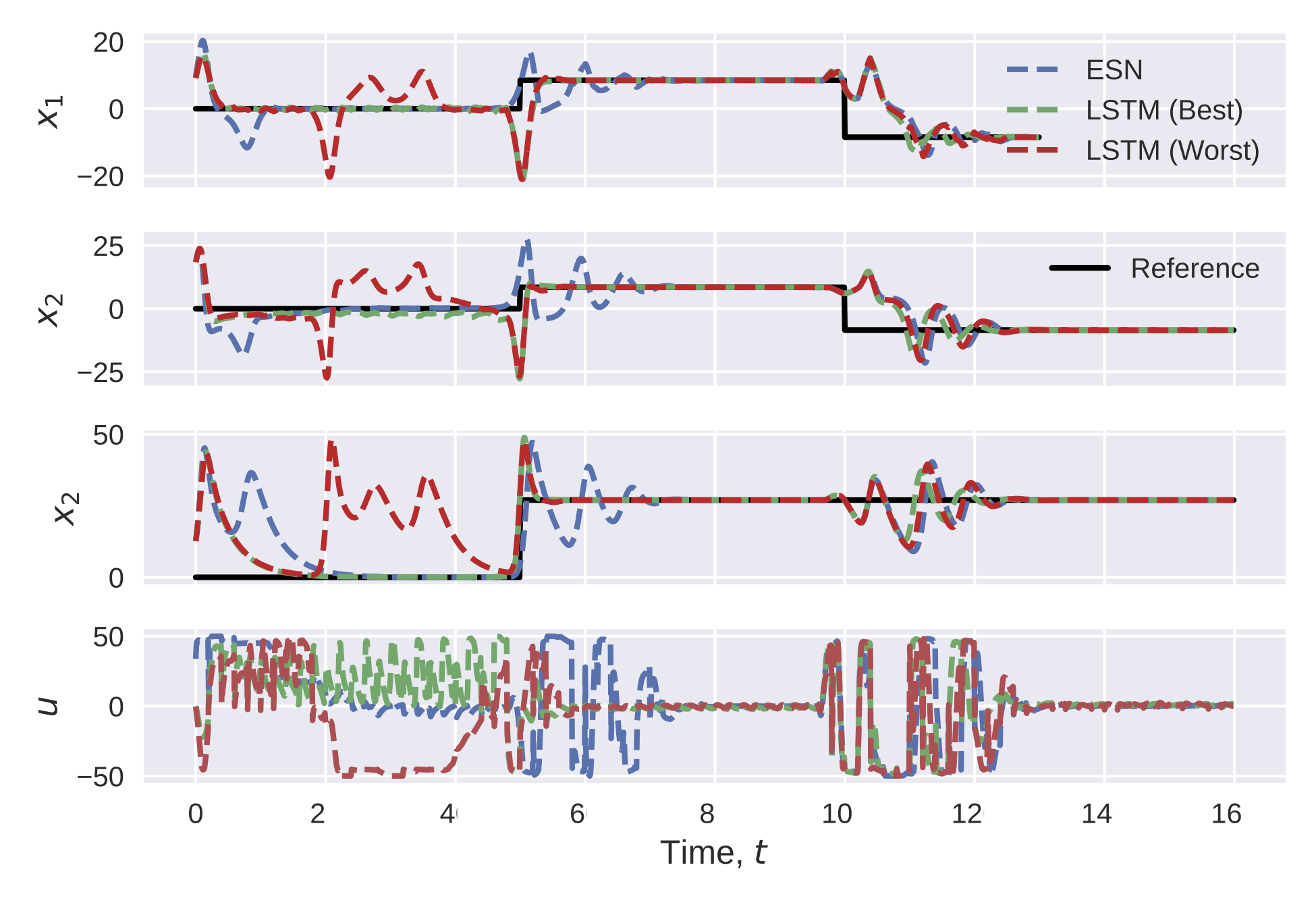} 
    \end{overpic}
    \caption{Control performance of LSTM and ESN based MPC in the Lorenz 63 system with control. Both the best and worst performing instances of LSTM-MPC are shown, while only the worst instance of ESN-MPC is illustrated. The aim of each controller is to successively steer an arbitrary initial condition to the fixed points at $(0,0,0)$, $(\sqrt{72}, \sqrt{72}, 27)$, and $(-\sqrt{72}, -\sqrt{72}, 27)$.}
    \label{fig:lorenz_control}
\end{figure*}

\begin{figure*}[!htb]
    \centering
    \begin{overpic}[width=0.77\textwidth]{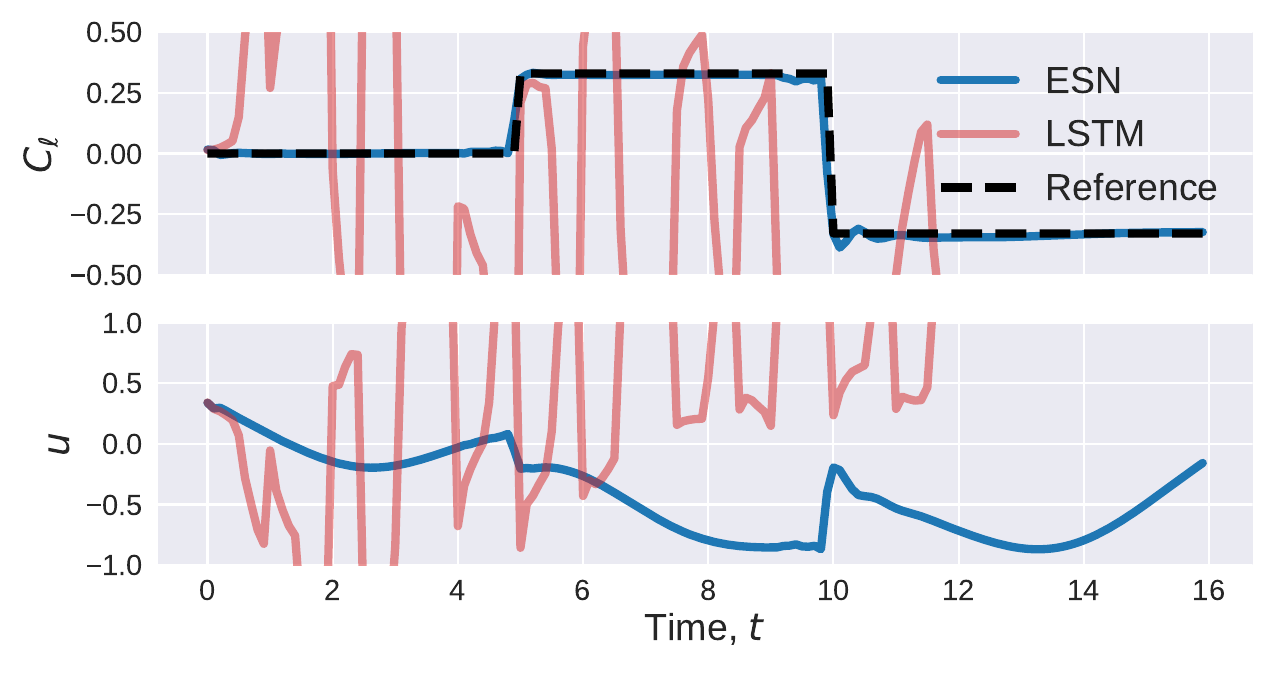} 
    \end{overpic}
    \caption{Control performance of LSTM and ESN based MPC for flow past the cylinder. The aim of each controller is to successively steer an arbitrary initial condition to coefficients of lift of 0, 0.33, and -0.33. The final four seconds of the LSTM trajectory has been truncated for clarity of presentation.}
    \label{fig:cyl_control}
\end{figure*}
To clarify the presentation and because the results of the previous section consistently show that ESNs and LSTMs yield the best forecasts, we present only the closed loop performance of LSTM and ESN based MPC in this section (color scheme has been changed to more clearly show distinction of ESN and LSTM performance).

\subsection{Linear, spring-mass system}
For the spring mass system, each MPC is tasked with successively steering the system to $x_1, x_2 = -1.5$ for 50 seconds, $x_1, x_2 = 0$ for 50 seconds, and $x_1, x_2 = 1.5$ for 50 seconds. Results for a particular instance of LSTM and ESN based MPC is shown in Fig. \ref{fig:lorenz_control}. Both controllers result in nearly identical control actions that successfully steer the system to the desired reference trajectory. The controllers also appear to correctly identify the steady state inputs required to maintain the system at each set point. To provide a more sensitive metric for comparing ESN and LSTM based MPC, we can compute $J(U,t)$ for each trajectory. Table \ref{tab:spring_mass_control} shows the mean and standard deviation of $J$ from 32 instances of both ESN and LSTM based MPC. The excellent performance of MPC with both surrogate models is to be expected in this case, as the underlying dynamics of the system are linear. 
\begin{table}[!htb]
    \centering
    \begin{tabular}{|c|c|c|}
    \hline
        Model & Mean $J$ & STD $J$\\
        \hline
        LSTM & 158 & 4\\
        \hline
        ESN & 159 & 1\\
        \hline
    \end{tabular}
    \caption{Mean and standard deviation of $J$ across 32 instances of ESN and LSTM based MPC for the spring mass system. }
    \label{tab:spring_mass_control}
\end{table}

\subsection{Continuous stirred tank reactor}
In contrast to the previous example where all set points were achievable, here the selected reference trajectory contains only one set point that is feasible (the first set point in Fig. \ref{fig:stirred_tank_control}). Both LSTM and ESN based MPC rapidly steer the system to this first set point and correctly identify the required constant control input to maintain the equilibrium. As shown in Fig. \ref{fig:stirred_tank_control}, both MPC algorithms yield similar control signals that strike a balance between achieving the set point in $x_1$ and $x_2.$ The mean and standard deviation across 32 instances of ESN and LSTM based MPC are displayed in Table \ref{tab:stirred_tank_control}. 
\begin{table}[!htb]
    \centering
    \begin{tabular}{|c|c|c|}
    \hline
        Model & Mean $J$ & STD $J$\\
        \hline
        LSTM & 326 & 0.3\\
        \hline
        ESN & 326 & 0.2 \\
        \hline
    \end{tabular}
    \caption{Mean and standard deviation of $J$ across 32 instances of ESN and LSTM based MPC for the stirred tank system. }
    \label{tab:stirred_tank_control}
\end{table}

\subsection{Two-tank reservoir}
In the case of the two-tank reservoir system, MPC must solve for two actuations at each time-step as opposed to only one. Fig. \ref{fig:two_tank_control} illustrates the control actions and state evolution of a particular instance of ESN and LSTM based MPC. In both cases, MPC achieves excellent performance following the desired reference trajectory. However, ESN based MPC appears to be more aggressive in the selected control actions when the set point switches between the three selected values. As evidenced in Table \ref{tab:two_tank_control}, ESN based MPC consistently outperforms LSTM based MPC and does so with significantly less variance in the cost, $J$.

\begin{table}[!htb]
    \centering
    \begin{tabular}{|c|c|c|}
    \hline
        Model & Mean $J$ & STD $J$\\
        \hline
        LSTM & 399 & 35 \\
        \hline
        ESN & 363 & 5\\
        \hline
    \end{tabular}
    \caption{Mean and standard deviation of $J$ across 32 instances of ESN and LSTM based MPC for the two-tank system. }
    \label{tab:two_tank_control}
\end{table}

\subsection{Lorenz Control}
The control task in the Lorenz system is to successively steer between the three fixed points of the system. These fixed points can be shown to exist at $(0,0,0)$, $(\sqrt{72}, \sqrt{72}, 27)$, and $(-\sqrt{72}, -\sqrt{72}, 27)$ for the selected values of $\sigma$, $\rho$, and $\beta$. Previous work has demonstrated that data-driven MPC methods can successfully steer the system to the latter two of these fixed points, however stabilizing the system at the origin is a significantly more challenging task. This difficulty arises for at least two reasons. First, the maximal real part of an eigenvalue associated with the Jacobian evaluated at the origin is larger than the maximal real part of eigenvalues of the Jacobian evaluated at the other two fixed points. Second, setting aside any initial transient, the training data never explores the phase space of the system near the origin. Consequently, we find that while all considered LSTM and ESN based controllers stabilize the system at $(\pm \sqrt{72}, \pm \sqrt{72}, 27)$, not all LSTM based MPC schemes sucessfully control the system to the origin. 

Table \ref{tab:lorenz_control} shows the mean and standard deviation of the MPC cost $J$ across 32 trials of ESN and LSTM based MPC. Fig. \ref{fig:lorenz_control} illustrates the best and worst performing instances of LSTM-MPC, as well as the worst performing instance of ESN-MPC. Not only does the worst case of LSTM based MPC fail to stabilize at the origin, even the best case of LSTM-MPC requires aggressive control actions to maintain the state at the origin. The worst case of ESN based MPC, on the other hand, still successfully steers the system to the origin and maintains the unstable set point with minimal control input. All models are able to achieve and maintain the other two set points.

\begin{table}[!htb]
    \centering
    \begin{tabular}{|c|c|c|}
    \hline
        Model & Mean $J$ & STD $J$\\
        \hline
        LSTM & 1050 & 149\\
        \hline
        ESN & 1042 & 11\\
        \hline
    \end{tabular}
    \caption{Mean and standard deviation of $J$ across 32 instances of ESN and LSTM based MPC for the Lorenz system with control. }
    \label{tab:lorenz_control}
\end{table}

\subsection{Flow past a cylinder}
The control objective for the flow past a cylinder is to control the coefficient of lift $C_{\ell}$ to 0, 0.33, and -0.33. Results for an ESN and an LSTM based MPC is shown in Fig. \ref{fig:cyl_control}. 
The LSTM based MPC entirely fails to control the system, while ESN based MPC provides excellent reference tracking. The work of Bieker et al. \cite{bieker_deep_2020} demonstrated that RNNs trained with gradient descent can be used to control the same system, but the amount of training data used in that work was much greater. This finding is consistent with our observations in Section IV, where we found that the ESN required less training data to achieve satisfactory forecasting performance.

\section{Discussion and Conclusion}
In this work, we demonstrated empirically that ESNs outperform other considered surrogate models (DMDc, FCNs, LSTMs, and GRUs) across all considered example systems (linear spring-mass, stirred tank reaction, coupled reservoir/tank system, Lorenz system with control, and flow past a cylinder) in the tasks of both forecasting and control. We believe the superior performance of ESN based surrogate models is due to two key reasons. First, ESNs avoid the issue of backpropagation through time  because ESN weights are computed using only a random initialization and linear regression. The rapid training that this allows yields significantly more efficient and effective hyperparameter sweeps in comparison to networks relying on gradient descent based learning. Second, the linear regression used to train ESNs requires fewer training samples than gradient descent based methods. As an example, in the Lorenz system with control, LSTMs required an order of magnitude more training data to achieve comparable forecasting performance. 

The combination of these two advantages allowed ESN based MPC to successfully control the lift generated by the flow past a cylinder, while assuming access to a fraction of the training data considered in previous work \cite{bieker_deep_2020}. No other architecture was successful at this task. Similarly, ESN based MPC consistently controlled the chaotic Lorenz system to a fixed point in a region of phase space that was not explored by the training data. These results support the conclusion that ESNs are better at extrapolating to unseen control and measurement inputs than the other architectures. Future work could further extend these advantages by leveraging recursive least squares algorithms to achieve rapid online updates to the ESN model. Online learning for gradient descent based methods is more difficult, but not impossible \cite{bieker_deep_2020}. Finally, there remain many unexplored candidate architectures for data-driven control. Variants of the S4 model offer a particularly attractive direction because of their strong foundation in control theory.

\section{Acknowledgments}

The authors acknowledge support from the National Science Foundation AI Institute in Dynamic Systems (grant number 2112085). The authors also thank Anastasia Bizyaeva and Dima Tretiak for insightful conversations.

\bibliography{rc-control.bib}

\begin{thebibliography}{10}

\bibitem{kouvaritakis_model_nodate}
B.~Kouvaritakis and M.~Cannon, {\em Model {Predictive} {Control}: {Classical}, {Robust}, and {Stochastic}}.
\newblock Advanced {Textbooks} in {Control} and {Signal} {Processing}, Springer Cham, 2016.

\bibitem{schwenzer_review_2021}
M.~Schwenzer, M.~Ay, T.~Bergs, and D.~Abel, ``Review on model predictive control: an engineering perspective,'' {\em The International Journal of Advanced Manufacturing Technology}, vol.~117, pp.~1327--1349, Nov. 2021.

\bibitem{wang_fully_2006}
J.-S. Wang and Y.-P. Chen, ``A fully automated recurrent neural network for unknown dynamic system identification and control,'' {\em IEEE Transactions on Circuits and Systems I: Regular Papers}, vol.~53, pp.~1363--1372, June 2006.
\newblock Conference Name: IEEE Transactions on Circuits and Systems I: Regular Papers.

\bibitem{morton_deep_2018}
J.~Morton, F.~D. Witherden, A.~Jameson, and M.~J. Kochenderfer, ``Deep dynamical modeling and control of unsteady fluid flows,'' in {\em Proceedings of the 32nd {International} {Conference} on {Neural} {Information} {Processing} {Systems}}, {NIPS}'18, (Red Hook, NY, USA), pp.~9278--9288, Curran Associates Inc., Dec. 2018.

\bibitem{bieker_deep_2020}
K.~Bieker, S.~Peitz, S.~L. Brunton, J.~N. Kutz, and M.~Dellnitz, ``Deep model predictive flow control with limited sensor data and online learning,'' {\em Theoretical and Computational Fluid Dynamics}, vol.~34, pp.~577--591, Aug. 2020.

\bibitem{lee_identification_2000}
C.-H. Lee and C.-C. Teng, ``Identification and control of dynamic systems using recurrent fuzzy neural networks,'' {\em IEEE Transactions on Fuzzy Systems}, vol.~8, pp.~349--366, Aug. 2000.
\newblock Conference Name: IEEE Transactions on Fuzzy Systems.

\bibitem{terzi_learning_2021}
E.~Terzi, F.~Bonassi, M.~Farina, and R.~Scattolini, ``Learning model predictive control with long short-term memory networks,'' {\em International Journal of Robust and Nonlinear Control}, vol.~31, no.~18, pp.~8877--8896, 2021.
\newblock \_eprint: https://onlinelibrary.wiley.com/doi/pdf/10.1002/rnc.5519.

\bibitem{jeon_lstm-based_2021}
B.-K. Jeon and E.-J. Kim, ``{LSTM}-{Based} {Model} {Predictive} {Control} for {Optimal} {Temperature} {Set}-{Point} {Planning},'' {\em Sustainability}, vol.~13, p.~894, Jan. 2021.
\newblock Number: 2 Publisher: Multidisciplinary Digital Publishing Institute.

\bibitem{huang_lstm-mpc_2023}
K.~Huang, K.~Wei, F.~Li, C.~Yang, and W.~Gui, ``{LSTM}-{MPC}: {A} {Deep} {Learning} {Based} {Predictive} {Control} {Method} for {Multimode} {Process} {Control},'' {\em IEEE Transactions on Industrial Electronics}, vol.~70, pp.~11544--11554, Nov. 2023.
\newblock Conference Name: IEEE Transactions on Industrial Electronics.

\bibitem{pan_model_2012}
Y.~Pan and J.~Wang, ``Model {Predictive} {Control} of {Unknown} {Nonlinear} {Dynamical} {Systems} {Based} on {Recurrent} {Neural} {Networks},'' {\em IEEE Transactions on Industrial Electronics}, vol.~59, pp.~3089--3101, Aug. 2012.
\newblock Conference Name: IEEE Transactions on Industrial Electronics.

\bibitem{draeger_model_1995}
A.~Draeger, S.~Engell, and H.~Ranke, ``Model predictive control using neural networks,'' {\em IEEE Control Systems Magazine}, vol.~15, pp.~61--66, Oct. 1995.
\newblock Conference Name: IEEE Control Systems Magazine.

\bibitem{jordanou_nonlinear_2018}
J.~P. Jordanou, E.~Camponogara, E.~A. Antonelo, and M.~A. Schmitz~Aguiar, ``Nonlinear {Model} {Predictive} {Control} of an {Oil} {Well} with {Echo} {State} {Networks},'' {\em IFAC-PapersOnLine}, vol.~51, pp.~13--18, Jan. 2018.

\bibitem{bonassi_nonlinear_2024}
F.~Bonassi, A.~La~Bella, M.~Farina, and R.~Scattolini, ``Nonlinear {MPC} design for incrementally {ISS} systems with application to {GRU} networks,'' {\em Automatica}, vol.~159, p.~111381, Jan. 2024.

\bibitem{bonassi_recurrent_2022}
F.~Bonassi, M.~Farina, J.~Xie, and R.~Scattolini, ``On {Recurrent} {Neural} {Networks} for learning-based control: {Recent} results and ideas for future developments,'' {\em Journal of Process Control}, vol.~114, pp.~92--104, June 2022.

\bibitem{kaiser_sparse_2018}
E.~Kaiser, J.~N. Kutz, and S.~L. Brunton, ``Sparse identification of nonlinear dynamics for model predictive control in the low-data limit,'' {\em Proceedings of the Royal Society A: Mathematical, Physical and Engineering Sciences}, vol.~474, p.~20180335, Nov. 2018.
\newblock Publisher: Royal Society.

\bibitem{rumelhart_learning_1987}
D.~E. Rumelhart and J.~L. McClelland, ``Learning {Internal} {Representations} by {Error} {Propagation},'' in {\em Parallel {Distributed} {Processing}: {Explorations} in the {Microstructure} of {Cognition}: {Foundations}}, pp.~318--362, MIT Press, 1987.
\newblock Conference Name: Parallel Distributed Processing: Explorations in the Microstructure of Cognition: Foundations.

\bibitem{hochreiter_long_1997}
S.~Hochreiter and J.~Schmidhuber, ``Long {Short}-{Term} {Memory},'' {\em Neural Comput.}, vol.~9, pp.~1735--1780, Nov. 1997.

\bibitem{hewamalage_recurrent_2021}
H.~Hewamalage, C.~Bergmeir, and K.~Bandara, ``Recurrent {Neural} {Networks} for {Time} {Series} {Forecasting}: {Current} status and future directions,'' {\em International Journal of Forecasting}, vol.~37, pp.~388--427, Jan. 2021.

\bibitem{platt_systematic_2022}
J.~A. Platt, S.~G. Penny, T.~A. Smith, T.-C. Chen, and H.~D.~I. Abarbanel, ``A systematic exploration of reservoir computing for forecasting complex spatiotemporal dynamics,'' {\em Neural Networks}, vol.~153, pp.~530--552, Sept. 2022.

\bibitem{vlachas_data-driven_2018}
P.~R. Vlachas, W.~Byeon, Z.~Y. Wan, T.~P. Sapsis, and P.~Koumoutsakos, ``Data-driven forecasting of high-dimensional chaotic systems with long short-term memory networks,'' {\em Proceedings of the Royal Society A: Mathematical, Physical and Engineering Sciences}, vol.~474, p.~20170844, May 2018.
\newblock Publisher: Royal Society.

\bibitem{chattopadhyay_data-driven_2020}
A.~Chattopadhyay, P.~Hassanzadeh, and D.~Subramanian, ``Data-driven predictions of a multiscale {Lorenz} 96 chaotic system using machine-learning methods: reservoir computing, artificial neural network, and long short-term memory network,'' {\em Nonlinear Processes in Geophysics}, vol.~27, pp.~373--389, July 2020.
\newblock Publisher: Copernicus GmbH.

\bibitem{vlachas_backpropagation_2020}
P.~R. Vlachas, J.~Pathak, B.~R. Hunt, T.~P. Sapsis, M.~Girvan, E.~Ott, and P.~Koumoutsakos, ``Backpropagation algorithms and {Reservoir} {Computing} in {Recurrent} {Neural} {Networks} for the forecasting of complex spatiotemporal dynamics,'' {\em Neural Networks}, vol.~126, pp.~191--217, June 2020.

\bibitem{armenio_model_2019}
L.~B. Armenio, E.~Terzi, M.~Farina, and R.~Scattolini, ``Model {Predictive} {Control} {Design} for {Dynamical} {Systems} {Learned} by {Echo} {State} {Networks},'' {\em IEEE Control Systems Letters}, vol.~3, pp.~1044--1049, Oct. 2019.
\newblock Conference Name: IEEE Control Systems Letters.

\bibitem{proctor_dynamic_2016}
J.~L. Proctor, S.~L. Brunton, and J.~N. Kutz, ``Dynamic {Mode} {Decomposition} with {Control},'' {\em SIAM Journal on Applied Dynamical Systems}, vol.~15, pp.~142--161, Jan. 2016.
\newblock Publisher: Society for Industrial and Applied Mathematics.

\bibitem{chung_empirical_2014}
J.~Chung, C.~Gulcehre, K.~Cho, and Y.~Bengio, ``Empirical {Evaluation} of {Gated} {Recurrent} {Neural} {Networks} on {Sequence} {Modeling},'' Dec. 2014.
\newblock arXiv:1412.3555 [cs].

\bibitem{maass_computational_2004}
W.~Maass and H.~Markram, ``On the computational power of circuits of spiking neurons,'' {\em Journal of Computer and System Sciences}, vol.~69, pp.~593--616, Dec. 2004.

\bibitem{jaeger_echo_no_date}
H.~Jaeger, ``"the ‘echo state’ approach to analyzing and training recurrent neural networks",'' tech. rep., German National Research Center for Information Technology, Technical Report GMD 148, 2001.

\bibitem{ames_control_2019}
A.~D. Ames, S.~Coogan, M.~Egerstedt, G.~Notomista, K.~Sreenath, and P.~Tabuada, ``Control {Barrier} {Functions}: {Theory} and {Applications},'' in {\em 2019 18th {European} {Control} {Conference} ({ECC})}, pp.~3420--3431, June 2019.

\end{thebibliography}
\bibliographystyle{ieeetr}

\end{document}